\documentclass[aps,
		prl,
		reprint,
        superscriptaddress,
        shortbibliography,
		  footinbib,
		floatfix,
		notitlepage, twocolumn
		]{revtex4-1}
\pdfoutput=1
\usepackage{amsmath}
\usepackage{mathtools}
\usepackage{amssymb}
\usepackage{slashed}
\usepackage{color}
\usepackage{xcolor}
\usepackage{hyperref}
\usepackage{braket}
\hypersetup{colorlinks=true,linkcolor=blue} 
\usepackage{comment}
\usepackage{bm}
\newcommand{\be}{\begin{equation}}
\newcommand{\ee}{\end{equation}}
\newcommand{\bi}{\begin{enumerate}}
\newcommand{\ei}{\end{enumerate}}
\newcommand{\ud}{{\mathrm{d}}}
\hfuzz=\maxdimen \tolerance=10000
\vfuzz=\maxdimen \tolerance=10000
\newcommand{\LCm}{{\scriptscriptstyle -}}
\newcommand{\LCp}{{\scriptscriptstyle +}}

\def\Put(#1,#2)#3{\leavevmode\makebox(0,0){\put(#1,#2){#3}}}
\newcommand{\vv}{{\rm v}}

\newcommand{\pd}{\partial}

\newcommand{\sv}{$\sqrt{\mathrm{Vaidya}}$}

\begin{document}

\title{Hawking radiation from the double copy}
\author{Anton Ilderton}
\email{anton.ilderton@ed.ac.uk}
\author{William Lindved} \email{william.lindved@ed.ac.uk}
\author{Karthik Rajeev}
\email{karthik.rajeev@ed.ac.uk}
\affiliation{Higgs Centre, School of Physics and Astronomy, University of Edinburgh, UK}

\begin{abstract}
Gravity and gauge theory are concretely linked by the double copy. 
Although well-studied at the level of perturbative scattering in vacuum, far less is known about non-perturbative aspects or extensions of double copy beyond trivial backgrounds. 
We show here how Hawking radiation in a collapse metric, its associated thermal spectrum, and horizon-dependence, emerges from the double copy of particle production in a background gauge field, where there is no global horizon, nor a thermal spectrum.
Our approach combines worldline and amplitudes methods, and allows the unification of several classical and quantum double copy prescriptions for black hole spacetimes. 
\end{abstract}

\maketitle
\paragraph{Introduction.} Double copy directly relates scattering amplitudes in a web of gauge and gravitational theories. This offers  both new insights into gravity and a practical tool for simplifying gravitational scattering calculations. For reviews see~\cite{Bern:2019prr,Borsten:2020bgv,Bern:2022wqg,Adamo:2022dcm}.  Double copy is typically applied at the level of perturbation theory, relating gluon and graviton scattering in Minkowski space, where it is well understood. Rather less is known about non-perturbative aspects of double copy~\cite{Monteiro:2011pc,Huang:2019cja,Alawadhi:2019urr,Banerjee:2019saj,Cheung:2022mix,Borsten:2021hua,Alawadhi:2021uie,Armstrong-Williams:2022apo} and double copy relations on background fields and/or curved spacetime~\cite{Luna:2016hge,Adamo:2017nia,Albayrak:2020fyp,Adamo:2020qru,Zhou:2021gnu,Diwakar:2021juk,Cheung:2022pdk,Herderschee:2022ntr,Drummond:2022dxd,Lee:2022fgr,Lipstein:2023pih,Mei:2023jkb,Liang:2023zxo,Brown:2023zxm,
CarrilloGonzalez:2024sto,Beetar:2024ptv,Ilderton:2024oly,Alday:2025bjp,Ilderton:2025gug}, both connected to the question of whether or not double copy is a fundamental property of gauge and gravitational interactions.

It is natural to ask, for example, how gauge theory can contain, or give rise to, gravitational structures which seemingly have no gauge theory equivalent. Black holes are an obvious target of study -- how is information on the horizon, boundary conditions, and Hawking radiation encoded in the gauge theory? It is known that the Schwarzschild metric can be obtained from the Coulomb potential through `classical double copy', which more broadly relates exact solutions of the classical Einstein and Maxwell equations~\cite{Monteiro:2014cda,Luna:2015paa,Berman:2018hwd,Alfonsi:2020lub,Bahjat-Abbas:2020cyb,Adamo:2021dfg,Luna:2022dxo,White:2024pve,Kim:2024dxo,Chawla:2024mse,Kent:2025pvu}, and within this scheme procedures exist for identifying horizons~\cite{Chawla:2023bsu}. What about the quantum theory, though? To what extent is information on black hole physics accessible through amplitude calculations?  We will address this question here.

Combining amplitudes-based approaches~\cite{Aoude:2023fdm,Aoude:2024sve,Aoki:2025ihc} with the worldline formalism~\cite{Affleck:1981bma,Bern:1990cu,Bern:1991aq,Strassler:1992zr,Edwards:2019eby,Kalin:2020mvi,Mogull:2020sak}, we will show here how Hawking radiation and associated structures emerge from double copy. The logic is simple. First, classical geodesics and conserved quantities in Coulomb and Schwarzschild backgrounds obey double copy relations~\cite{Gonzo:2021drq}. Second, scattering amplitudes in the semiclassical regime can be calculated via worldline methods which require only knowledge of classical geodesics~\cite{Hartle:1976tp,Chitre:1977ip,Srinivasan:1998ty,Semren:2025dix,Ilderton:2025umd}. This suggests that Hawking radiation should emerge, at all orders in the coupling, by applying what are essentially classical double copy relations
to worldline expressions for gauge theory particle creation amplitudes.
Focussing on the case of massless particles, we will show that Hawking radiation in a Vaidya spacetime is indeed the double copy of pair creation from a gauge field which is related to Vaidya by classical notions of double copy.
As part of this we will also give an exact calculation of particle creation on the gauge theory side, identifying structures which are natural `single copy' analogues of those in gravity, and in existing calculations of Hawking radiation.

Beginning in Yang-Mills, we write the Minkowski metric in ingoing coordinates as
\begin{align}
   \eta_{\mu\nu}\ud x^\mu \ud x^\nu = \ud v^2 -2\ud v\,\ud r-r^2\ud \Omega^2 \;,
\end{align}
and define the null vector $k_{\mu}$ through $k_{\mu}\ud x^{\mu}=\ud v$.
In these coordinates our chosen Yang-Mills potential is
\begin{align}\label{YMfield}
    A^a_{\mu}=\frac{g\tilde{c}^{a}}{r}\Theta(v)k_{\mu} \;,
\end{align}
for coupling $g$ and colour factors ${\tilde c}^a$. This Coulomb-like potential, which forms at $v=0$, models the fields of a charged, radially collapsing null shell (see Appendix~\hyperref[appA]{A} for details).
In complete analogy to the case of Coulomb and Schwarzschild~\cite{Monteiro:2014cda}, we generate a metric perturbation $h_{\mu\nu}$ from (\ref{YMfield}) by applying (in our conventions) the standard Kerr-Schild replacement rules
\be\label{DCrule1}
    g^2 \to 2 G \,, \qquad {\tilde c}^a \to M k_\mu \;,
\ee
where $G$ is Newton's constant. The full metric, $g_{\mu\nu} =\eta_{\mu\nu} - \sqrt{2G} h_{\mu\nu}$, is then the Vaidya spacetime
\be\label{Vaidya-metric}
    g_{\mu\nu} = \eta_{\mu\nu} - \frac{2GM}{r}\Theta(v)k_\mu k_\nu \;,
\ee
describing the formation of a black hole, of mass $M$, from the radial collapse of a 
spherical null shell~\cite{Vaidya:1966zza}.
In analogy to the literature on e.g.~the single copy of Kerr~\cite{Monteiro:2014cda,Arkani-Hamed:2019ymq,Guevara:2020xjx}, we will refer to the gauge field (\ref{YMfield}) as \sv.
\begin{figure}
    \includegraphics[width=0.49\linewidth]{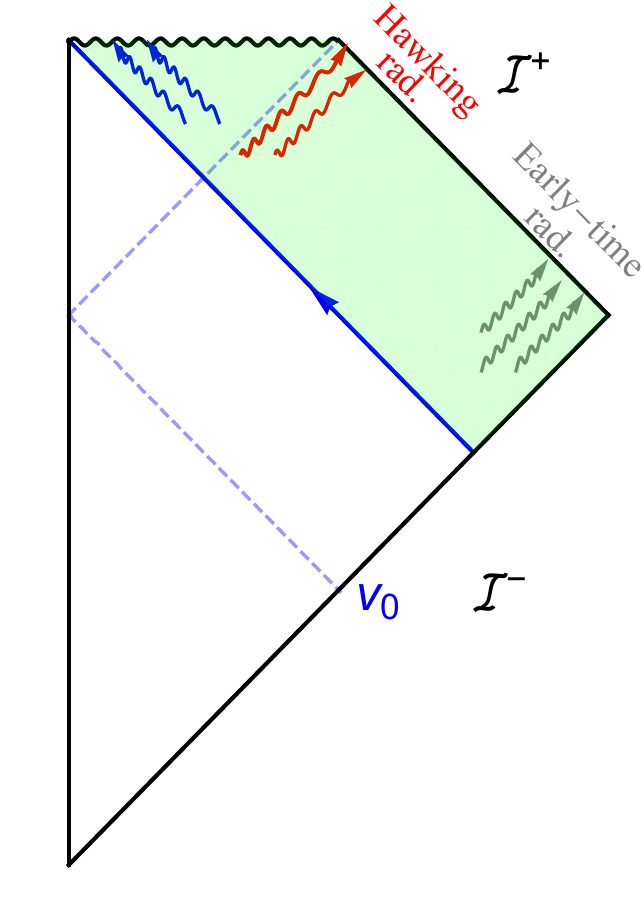}
  \hfill 
    \includegraphics[width=0.41\linewidth]{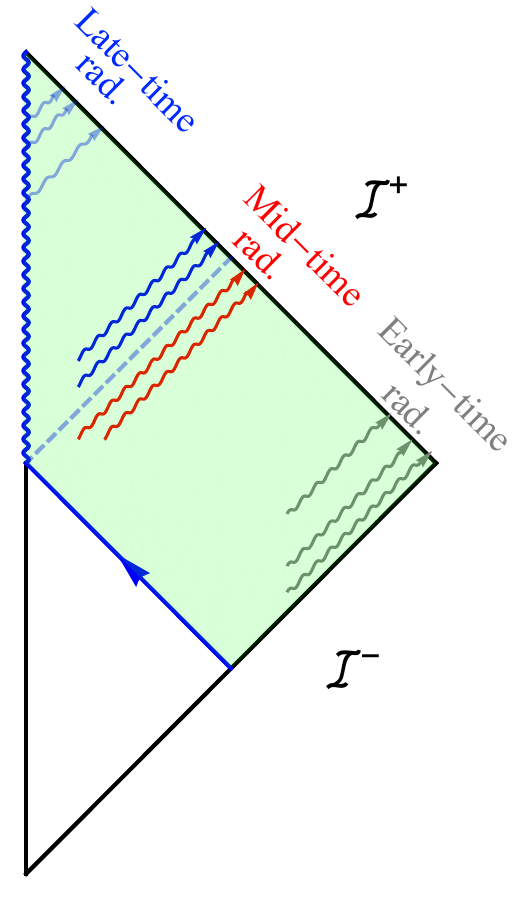}
\caption{    \label{fig:time_scales}
 Left: Penrose diagram for the Vaidya spacetime (\ref{Vaidya-metric}) in which a null shell collapses to form a black hole. The shaded region is described by the Schwarzschild metric. Asymptotic regions of early-time and Hawking radiation are illustrated. Right: the single copy \sv, that is, flat Minkowski space in which a Coulomb field forms at $v=0$  (shaded region). Emitted radiation is illustrated for early, late, and mid (or `Hawking-like') times.}
\end{figure}
\paragraph{From classical to quantum.}
Classical geodesics in a Coulomb field are fully determined by their conserved quantities. These double copy, using (\ref{DCrule1}) and replacing certain colour factors with kinematic factors,  to the corresponding conserved quantities for geodesics in the Schwarzschild metric, at the level of the equations of motion~\cite{Gonzo:2021drq}. Different prescriptions exist in different scenarios, though, see e.g.~\cite{Goldberger:2016iau,Adamo:2020qru,Moynihan:2025vcs}, so our first task will be to identify the double copy rules appropriate to \emph{solutions} of the equations of motion in Vaidya and \sv.

For a massless particle, trajectory {$z^\mu(\lambda)$} and charge $g c^a$, the colour factor $c^a {\tilde c}^a$ is conserved in the background (\ref{YMfield}), 
so it is enough to consider the Abelian problem~\cite{Gonzo:2021drq}. We therefore write $c^{a}\tilde{c}^a=-q Q$ from here on, in which  $q$ and $Q$ are (positive) `atomic numbers' of the Abelian probe particle~\footnote{Our conventions are such that what we call a `particle' is attracted to the source, in both Yang Mills and gravity.} and background, respectively. The sign in $c^{a}\tilde{c}^a$ means that the resulting interaction is attractive.
The relevant action for a massless particle in {\sv} is then
\begin{align}\label{eq:def_action}
    S[z]= \int^{\lambda_i}_{\lambda_f}\mathcal{L}\, d\lambda\quad;\quad 
    \mathcal{L}=
    \tfrac{1}{2}\dot{z}\cdot\dot{z}
    +a\cdot \dot{z}\;,    
\end{align}
with effective Abelian potential $a_{\mu}=-(g^2qQ/r)\Theta(v)k_{\mu}$. Turning to the equations of motion, the energy $\partial\mathcal{L}/{\partial\dot{v}}$ is separately conserved in the vacuum and Coulomb regions. However, the time-dependence of the background means that the energy changes as particles cross the null ray $v=0$. Solving the equations of motion for a probe particle, initially radially infalling with energy $\mathcal{E}$, the change in energy $\Delta \mathcal{E}$ as the particle crosses $v=0$ into the Coulomb region is
\begin{align}\label{energies-root-vaidya}
    \Delta\mathcal{E}=-\left.\frac{g^2qQ}{r}\right\rvert_{v\rightarrow 0^{-}}=\frac{2g^2qQ}{\vv} \;,
\end{align}
where ${\vv}<0$ is the $v$-coordinate on $\mathcal{I}^-$ at which the geodesic originates. Now let $p_\mu=\delta\mathcal{L}/\delta \dot{z}^\mu$ be the probe momentum and $P=M d(v-r)$ the momentum of the black hole's source; we observe, in the spirit of~\cite{Goldberger:2016iau,Gonzo:2021drq}, that the double copy rules $qQ\rightarrow P\cdot p$ and $g^2\rightarrow 2G$ correctly map~(\ref{energies-root-vaidya}) to the analogous energy difference in the Vaidya metric, for a radial null geodesic travelling from the vacuum to the Schwarzschild region (as can be verified by direct calculation~\cite{Ilderton:2025umd}):
\be\label{energies-vaidya}
    \Delta\mathcal{E} 
    =-\left.\frac{2G\,P\cdot p}{r}\right\rvert_{v\rightarrow 0^-}=
    \frac{4GM \mathcal{E}}{\vv}
    \;,
\ee
where ${\vv}$ is again the initial coordinate of the null geodesic at $\mathcal{I}^\LCm$. Comparing (\ref{energies-root-vaidya}) and (\ref{energies-vaidya}) suggests that at the level of the \emph{solutions} of the equation of motion, we have the simple double copy map $qQ\rightarrow M\mathcal{E}$. However, when we discuss scattering amplitudes, we will find it convenient to express quantities in terms of the \emph{outgoing} energy $\mathcal{E}':=\mathcal{E}+\Delta \mathcal{E}$. To this end we observe that the double copy map could equivalently be stated as 
\be\label{new-DC-rule-1}
    {qQ\to M\mathcal{E}'
    \quad
    \&
    \quad 
    {\vv} \to {\vv}-{\vv}_0 \;,}
\ee
where ${\vv}_0:=-4GM$; {this yields the correct energy difference (\ref{energies-vaidya}) in Vaidya.} Sending ${\vv\to\vv}-{\vv}_0$ is motivated by observing that ${\vv}_0$ is the initial $v$--coordinate for the last radial null geodesic that escapes the collapse in Vaidya, while the corresponding point in \sv\, is $\vv=0$. 
We will confirm below {that (\ref{new-DC-rule-1}) is indeed the rule obeyed at the \emph{amplitude} level.}

As a first step, we consider the 1-to-1 amplitude for a massless scalar scattering on (\ref{YMfield}) or (\ref{Vaidya-metric}). Following \cite{Aoude:2024sve}, we transform this amplitude to a suitably defined `impact parameter space' via the on-shell Fourier transform
\begin{align}
    i\tilde{\mathcal{A}}({\vv})=\int \frac{\ud^4q}{(2\pi)^3}\delta(p'^2-2p'\cdot q)\braket{p'|i\mathcal{T}|p'-q}e^{-ib\cdot q}
\end{align}
in which $\mathcal{T}$ is the transition matrix, $q$ is the momentum transfer and $b=({\vv},\mathbf{0})$ is the analogue of an impact parameter.
We then define eikonal phase $\mathcal{\chi}$ in the usual manner via (momentarily retaining $\hbar$)  $e^{i\frac{\chi({\vv})}{\hbar}}\left(1+\mathcal{O}(\hbar)\right)=1+i\tilde{\mathcal{A}}({\vv})$~\cite{DiVecchia:2023frv}.
Since variation of $\chi$ with respect to impact parameter leads to the impulse via $\partial_{\vv}\chi({\vv})=-\Delta\mathcal{E}$, we immediately infer $\chi({\vv})$ for {\sv} and Vaidya from \eqref{energies-root-vaidya} and \eqref{energies-vaidya} respectively.
{Since the eikonal phases in the two theories depend on $\mathcal{E}'$ and ${\vv}$, it seems sensible that they should be related by (\ref{new-DC-rule-1}).} 
To see this explicitly, we can consider the tree-level contribution to $\chi({\vv})$.
The calculations are analogous in Vaidya and \sv, and are discussed along with extensions in~\cite{Aoude:2024sve} and~\cite{Aoude:2025jvt}, so we simply quote the two results
\be\label{pert-comparison}
i\chi_{\rm tree}({\vv}) \simeq
\begin{cases}
    -i2g^2 qQ \log({-{\vv}}/{r_0}) & \text{\sv}, \\[2pt]
     -4iGM\mathcal{E}'\log({-{\vv}}/{r_0}) & \text{Vaidya}\,,
\end{cases} 
\ee
where $r_0$ is some scale. These expressions are consistent {with
(\ref{new-DC-rule-1}), but} seemingly only modulo the rule ${\vv\rightarrow \vv}-{\vv}_0$; however, (\ref{pert-comparison}) only holds to leading order in the coupling, and it has already been shown in~\cite{Aoude:2024sve} that by including subdominant corrections, the expected dependence of the Vaidya result on $\vv-{\vv}_0$ indeed re-emerges.

\paragraph{Hawking radiation from the double copy.}
We move on to massless scalar pair creation in {\sv}.
This process can be considered, as illustrated in Fig.~\ref{fig:time_scales}, at early, mid and late times. The early-time spectrum is analogous to the so-called pre-Hawking radiation in the gravitational case~\cite{Gerlach:1976ji,Vachaspati:2006ki}. 
In the limit that $g^2 qQ \gg 1$ the well-known instability of the Coulomb field drives late-time radiation, where particles of the same charge as the source dominate the outgoing spectrum. Of interest to us is the mid-time spectrum, which is dominated by highly red-shifted particles that have opposite sign to the source, hence interact attractively, as would be the case in gravity.  We will now show that the mid-time radiation in \sv$\,$ double-copies to Hawking radiation in Vaidya.

Given that the classical gauge problem is essentially Abelian, it is enough to consider the Klein-Gordon equation for a massless particle in the {\sv} background,
\begin{align}\label{eq:KG}
    (\partial_{\mu}+ia_{\mu})(\partial^{\mu}+ia^{\mu})\varphi=0 \,,
\end{align}
with, as before, potential $a_{\mu}=-(g^2qQ/r)\Theta(v)k_{\mu}$.
To connect, below, to conventional discussions of Hawking radiation, we consider `out-mode' solutions of (\ref{eq:KG}) which are essentially free at $\mathcal{I}^\LCp$. To connect to our classical discussion, we also restrict to zero-angular momentum modes. As such we seek solutions of \eqref{eq:KG} with the boundary condition
\begin{equation}\label{asymptotic-behaviour}
   \varphi^{\rm(out)}_{\mathcal{E}'}(x)\sim \frac{e^{- i \mathcal{E}' {(v-2r)}} e^{2 i g^2 qQ \ln \frac{r}{r_0}}}{r} \;,
\end{equation}
as $x\sim \mathcal{I}^\LCp$, where $\mathcal{E}'$ is again the outgoing energy. The exact forms of these solutions will be given shortly. However, to extract information on particle production, standard results on Bogoliubov transformations~\cite{Fradkin1991QED,Wald1994QFT} tell us that we only need the asymptotic form of $\varphi^{\rm(out)}_{\mathcal{E}'}(x)$ in the past, i.e.~as $x\sim \mathcal{I}^-$. This is conveniently extracted from the following (Lorentzian) worldline path-integral representation of the wavefunction~\cite{Feldbrugge:2019sew,Rajeev:2021zae}: 
\begin{align}\label{worldline-def}
 \varphi^{\rm(out)}_{\mathcal{E}'}(x) =\int_{-\infty}^{\infty}\!\ud T
 \int_{\mathrm D}^{\mathrm N}
 \!\mathcal{D}[z]e^{iS[z]}    \;,
\end{align}
where $z^{\mu}(\lambda)$ are the worldline coordinates, the action is as in \eqref{eq:def_action} and $T=\lambda_f-\lambda_i$ is the proper time. The Dirichlet (D) and Neumann (N) boundary conditions are
\begin{align}
    {\mathrm D:}\quad &  z^\mu(\lambda_i)=x^\mu\,, \\
    {\mathrm N:}\quad &
    \dot{u}(\lambda_f)=0\,,
    \quad
    \dot{r}(\lambda_f)-\tfrac{g^2qQ}{r(\lambda_f)}=\mathcal{E}'\,,
\end{align}
where $u=v-2r$, along with $\dot{\theta}(\lambda_f)=\dot{\phi}(\lambda_f)=0;$ the Neumann conditions simply impose radial-outgoing conditions at proper time $\lambda_f$. These mixed boundary conditions can be effected by adding appropriate boundary terms to the action, see~\cite{Ilderton:2025umd} for the Vaidya case. 
We will evaluate the integrals in (\ref{worldline-def}) to all orders in the coupling and in the semiclassical limit; in this saddle-point approximation the integrals return (a fluctuation factor times) the exponential of the worldline action evaluated on a solution of the equations of motion. But, for $x$ on $\mathcal{I}^{-}$ where we want to evaluate $\varphi^{\mathrm{(out)}}_{\mathcal{E}'}$,  this `saddle-point' worldline $z_{cl}$ is precisely the radially infalling geodesic discussed above, which has prescribed final energy $\mathcal{E}'$ and which originates from $\mathcal{I}^-$ at $v=\vv$.
Thus the wavefunction has the form
\be\label{eq:out-gen}
 \varphi^{\rm(out)}_{\mathcal{E}'}(x)\propto \frac{1}{r}\, e^{iS_{cl}(v)}\Theta(-v) \;,   
\ee
where $S_{cl}(v) = S[z_{cl}]$ is the classical action, and
the factor of $\Theta(-v)$ highlights that only worldlines with $\vv<0$ (those escaping the Coulomb potential to $\mathcal{I}^{+}$) contribute. $S_{cl}(v)$ can be computed explicitly, but a quick method of computation is to observe that it must obey $\partial_{v}S_{cl}(v)=-\mathcal{E}(v)= -\mathcal{E}'+\Delta \mathcal{E}$, which can be easily integrated to arrive at
\begin{align}\label{wavefunct-sv}
    \varphi^{\rm(out)}_{\mathcal{E}'}(x)\propto
    \frac{1}{r}
    e^{-i\mathcal{E}'v+2ig^2qQ\log(-v/r_0)}\Theta(-v)\,.
\end{align}
We immediately observe that {the double copy map} \eqref{new-DC-rule-1} yields, up to an irrelevant phase,
\begin{align}\label{wavefunct-v}
    \varphi^{\rm(out)}_{\mathcal{E}'}(x)\to \frac{1}{r} e^{-i\mathcal{E}'v+4iGM\mathcal{E}'\log\left(\frac{{\vv}_0-v}{r_0}\right)}\Theta({\vv}_0-v)\,.
\end{align}
which is the correct expression for the leading semiclassical approximation of the out-mode in the Vaidya spacetime~\cite{Ilderton:2025umd}. (It is also the universal form of the same mode near $v\sim \vv_0$ in a general radial collapse~\cite{Hawking:1975vcx}.) Thus the semiclassical wavefunctions in \sv\, and Vaidya are related by the same double copy map as the classical geodesics \footnote{{We note in passing that the prescription $qQ\rightarrow M\mathcal{E}$ mentioned ahead of~\eqref{new-DC-rule-1}}, enforced at the level of the \emph{action} by $qQ\rightarrow -M\partial_{v} S_{cl}(v)$ {where} $\partial_{v}S_{cl}(v)=-\mathcal{E}'+\Delta \mathcal{E}$, also gives the correct semiclassical action relevant to the mode function~(\ref{eq:out-gen}). It would be interesting to explore the connection between this procedure and the observation in~\cite{Adamo:2022rmp} that certain aspects of double copy in plane-wave backgrounds manifest when momentum conserving delta-functions on the gauge-theory side are represented in position-space form.}. We can now consider the double copy of amplitudes and observables by following standard procedures to extract the Bogoliubov coefficients from our wavefunctions. These are computed by (Klein-Gordon) overlaps between in- and out-modes evaluated on $\mathcal{I}^\LCm$~\cite{Adamo:2023cfp}:
\begin{equation}\label{KG-overlap}
    \mathcal{F}(\mathcal{E}', \pm \mathcal{E})\equiv \lim_{r\to\infty}\! \int_{-\infty}^\infty\!\! \ud v \, 8 \pi r^2 \, \varphi^{(\text{out})}_{\mathcal{E}'} \, i \pd_v 
    \Big(\frac{e^{\pm i \mathcal{E}v} }{r} \Big) \;,
\end{equation}
where we used the explicit form of free, definite energy, radial in-modes in the asymptotic past. In the same region the radial out-modes look like functions of $v$ divided by $r$ (this will be shown for exact solutions below, but is already clear at the semiclassical level in (\ref{wavefunct-sv}) and (\ref{wavefunct-v})), hence (\ref{KG-overlap})  effectively reduces to a Fourier transform.
For positive energy in-modes the positive sign in (\ref{KG-overlap}) should be taken, which returns Bogoliubov $\alpha(\mathcal{E}', \mathcal{E})$, whilst the overlap with negative energy in-modes gives $\beta(\mathcal{E}', \mathcal{E}) = \mathcal{F}(\mathcal{E}', -\mathcal{E})$.

As is standard, $|\beta|^2$ gives the number density of created particles, while $\alpha^{-1}\beta$ is the pair creation amplitude~\cite{Fradkin1991QED,Wald1994QFT}.
Calculating these overlaps for our \sv\, wavefunctions, assuming large redshift $\mathcal{E}'\ll\mathcal{E}$, we find
\begin{align}\label{amp-gauge}
   \left\lvert\alpha(\mathcal{E}',\mathcal{E})^{-1}\beta(\mathcal{E}',\mathcal{E})\right\rvert \sim e^{-2\pi g^2qQ} \;.
\end{align}
The double copy of this mid-time particle creation amplitude (\ref{amp-gauge}) is, {again using} (\ref{new-DC-rule-1}),
\be\label{amp-grav}
      \left\lvert\alpha(\mathcal{E}',\mathcal{E})^{-1}\beta(\mathcal{E}',\mathcal{E})\right\rvert  \to e^{-4\pi GM \mathcal{E}'} \;,
\ee
which is, as claimed, the Hawking radiation amplitude in Vaidya.
The associated particle spectrum is thermal~\cite{Hawking:1974rv}, even though the gauge theory result (\ref{amp-gauge}) is not. Note that, unlike for the wavefunctions, the rule $v\rightarrow v-{\vv}_0$ does not play a role in the double copy of the {modulus of the pair creation} amplitude (\ref{amp-gauge}), or the spectrum (it only introduces a phase factor $e^{i\mathcal{E}{\vv}_0}$ into the Bogoliubov coefficients). This is consistent of course, since using (\ref{wavefunct-v}) in (\ref{KG-overlap}) yields (\ref{amp-grav}) directly.

\paragraph{The exact calculation in gauge theory.}
As alluded to above, the particle spectrum in $\sqrt{\text{Vaidya}}$ can be computed exactly.  Doing so will illuminate how and in which precise limits the semiclassical result, and therefore the single copy of Hawking radiation, emerges.

To begin we need the exact solution to the Klein-Gordon equation (\ref{eq:KG}) to  use with (\ref{KG-overlap}). In the Coulomb region $v>0$, a radial mode function
has the form
\begin{equation}\label{ansatz}
    \varphi^{\rm (out)}_{\mathcal{E}'}(x) = \mathcal{N} e^{- i \mathcal{E}' (v-r) } e^{i g^2qQ \ln \frac{r}{r_0}} \frac{w(r)}{r} \;, 
\end{equation}
where $\mathcal{N} $ is a normalisation constant and $w(r)$ obeys
\begin{equation}
    w''(r) + \left( \mathcal{E}' + \frac{g^2qQ}{r} \right)^2 w(r) = 0. 
\end{equation}
This is solved by either of the two Whittaker functions $M_{\kappa, \mu}(2 i \mathcal{E}'r)$ or $W_{\kappa, \mu}(2 i \mathcal{E}'r)$, with $\kappa = - i g^2qQ$ and $\mu = i \sqrt{g^4q^2 Q^2 - 1/4}$~\cite{DLMF:Whittaker}, corresponding to two physically distinct modes; $M_{\kappa,\mu}$ describes a purely outgoing wave near the origin, which then propagates out to reach $\mathcal{I}^+$, while $W_{\kappa,\mu}$ contains both incoming and outgoing waves near the origin and  vanishes near $\mathcal{I}^+$, describing a mode that is trapped inside the potential. This behaviour is analogous to that of the future asymptotic modes in a black hole spacetime, which split into out-modes (those which have support on $\mathcal{I}^+$) and horizon-modes (which have no support on $\mathcal{I}^+$~\cite{Hawking:1975vcx}).

Constructing solutions to (\ref{eq:KG}) requires extending (\ref{ansatz}) across the $v=0$ boundary, beyond which it must satisfy the free Klein-Gordon equation. Free radial solutions at $v<0$ are, requiring continuity at the origin, of the form
\begin{equation}
\frac{f(v-2r) - f(v)}{r}
\end{equation}
where $f$ is arbitrary.  Continuity of $\varphi_{\mathcal{E}'}^{(\text{out})}$ at $v=0$ then fixes
\begin{equation}
    f(s) = \mathcal{N} e^{- i \frac{\mathcal{E}'}{2} s} e^{i g^2qQ \ln -\frac{s}{2r_0}} w \left( - \frac{s}{2}  \right) \;.
\end{equation}
We now focus on the modes starting at future null infinity, for which $w(r) = M_{\kappa, \mu}(2 i \mathcal{E}' r)$, and fix the normalisation constant $\mathcal{N}$ such that the mode function (\ref{ansatz}) complies with the outgoing boundary conditions (\ref{asymptotic-behaviour}).
Further, we implicitly assume the presence of a wavepacket
which localises the mode somewhere in the mid-time region of $\mathcal{I}^+$.  With this we are ready to compute the Bogoliubov coefficients of the mid-time modes, which as we discussed in the previous section are those which double copy to Hawking radiation. Tracing  the out-mode back toward $\mathcal{I}^-$, it behaves differently on either side of the $v=0$ boundary. In the Coulomb region, the mode function is proportional to $e^{- i \mathcal{E}'v}/r$, describing standard Coulomb scattering from a part of the wave that has been reflected. Consequently the integration region $v>0$  contributes with only a pole term to the Bogoliubov coefficients. 

The parts of the Bogoliubov coefficients that are relevant for particle production come from the region close to $\mathcal{I}^-$ before the collapse at $v=0$, where the mode function takes the form
\begin{equation}\label{eq:out_with_Whitt}
    \varphi^{\rm (out)}_{\mathcal{E}'}(x) \sim -\mathcal{N} \frac{e^{- i \frac{\mathcal{E}'}{2} v} e^{i g^2qQ \ln - \frac{v}{2r_0}}}{r} M_{\kappa, \mu}(- i \mathcal{E}' v) \;.
\end{equation}
As we are implicitly considering a wavepacket-mode localised in the mid-time region of $\mathcal{I}^+$, (\ref{eq:out_with_Whitt}) really describes the part of the wavepacket which has been transmitted through the potential barrier in the Coulomb region, to then enter the free region close to the singularity at $r = 0$. Tracing it back further, the wavepacket ends up at small negative $v$ on $\mathcal{I}^-$, where the mode function inherits the small $r$ behaviour in the Coulomb region through the patching procedure at $v = 0$. Moreover, the small $|v|$ expansion of the right-hand-side of \eqref{eq:out_with_Whitt} can easily be found to match precisely with that of~\eqref{wavefunct-sv}, hence justifying the semiclassical approximation used to obtain it. Thus these modes indeed probe the dynamics close to the Coulomb \emph{singularity}, which plays an analogous role to the black hole \emph{horizon} in Hawking radiation.

The Bogoliubov coefficients relevant to pair creation can now be computed by  using \eqref{eq:out_with_Whitt} in \eqref{KG-overlap}, which yields
\begin{align}
    \mathcal{F}&(\mathcal{E}', \pm\mathcal{E})\propto \hspace{-.2em} \int_{- \infty}^0 \!\hspace{-.2em} \ud v \, e^{- i \left( \frac{\mathcal{E}'}{2} \mp \mathcal{E} \right)v } \hspace{-.3em} \left(\! -\frac{v}{2r_0}\!\right)^{- \kappa} \hspace{-.3em} M_{\kappa, \mu}( - i \mathcal{E}' v)\;.\nonumber
\end{align}
This integral can be performed exactly in terms of hypergeometric functions, see Appendix~\hyperref[app:B]{B} for details. However, the expression is greatly simplified in the limit $|\mathcal{E}| \gg \mathcal{E}'$, i.e.~when there is severe redshift. This is precisely what we expect from the classical geodesics (\ref{energies-root-vaidya}) of these mid-time modes which end up at small negative $v$ on $\mathcal{I}^-$. The semiclassical limit corresponds (reinstating $\hbar$) to $g^2 qQ/\hbar\gg 1$, consistent with a macroscopic Coulomb charge. In this limit, we can therefore approximate $\mu \simeq i g^2 qQ/\hbar$. With these approximations in place the Bogoliubov coefficients satisfy the relationship (\ref{amp-gauge}), thus recovering the amplitude which double copies to that of Hawking radiation.

\paragraph{Conclusions.}
We have used a novel combination of worldline and amplitudes-based methods to show that Hawking radiation in a collapse metric is the double copy of pair creation in a background gauge field. The metric and gauge backgrounds are themselves related by classical Kerr-Schild double copy. We first worked semiclassically, but to all orders in $G$, showing that mid-time radiation (massless particle creation) in {\sv} double copies to Hawking radiation. The appropriate double copy map is inherited from that obeyed by classical geodesics and conserved quantities in the two spacetimes, and this is responsible for introducing the energy dependence which results in a thermal Hawking spectrum. Our results thus tie together \emph{three} notions of double copy in black hole physics; that for classical solutions, for geodesics, and for amplitudes. This makes it clear that, ultimately, the symmetries of the background, hence conserved quantities, are responsible for the double copy relations between the gauge and gravity theories. (See~\cite{Kent:2025pvu} for a discussion of classical double copy in terms of Killing vectors.)

Turning to future work, and referring to Fig.~\ref{fig:time_scales}, early-time radiation in the gauge theory seems to have a natural analogue on the gravitational side, but one could ask what happens to the \emph{late-time} radiation in \sv\, under double copy. Not unrelated, it would be interesting to repeat our exact wavefunction calculation in Vaidya, using the required Heun functions, or in $\sqrt{\text{Kerr}}$.

Another avenue to explore would be information-theoretic aspects of the single copy of Hawking radiation, through e.g.~the reduced density matrix of particle modes in \sv, and what implications this might have for the information-loss paradox and related topics. It would also be worth investigating the role of backreaction in the single copy problem, and to assess what insights its double copy provides into analogous aspects of Hawking radiation, where several open questions remain.

\begin{acknowledgments}
    \textit{We thank T.~Adamo for feedback on the manuscript, and R.~Aoude, D.~O'Connell, M.~Sergola and C.D.~White for discussions and for sharing a draft of~\cite{Aoude:2025jvt}. The authors are supported by the STFC Consolidated Grant ST/X000494/1 (AI) ``Particle Theory at the Higgs Centre" (AI, KR) and an STFC studentship (WL).}
\end{acknowledgments}

\appendix
\onecolumngrid
\section*{End Matter}

\paragraph{Appendix A:}\label{appA}
The Yang-Mills potential (\ref{YMfield}) models the field of a charged, radially collapsing null shell. It is sufficient to assume that the colour structure factorises, such that the field is essentially Abelian. Replacing ${\tilde c}^a \to Q$, the potential and field strength are then
\begin{align}
A = \frac{gQ}{r}\Theta(v)\, \ud v\;, 
\qquad
    F =\frac{gQ}{r^2}\Theta(v) \,\ud v\wedge \ud r\,.
\end{align}
The field is sourced by the current \;
\begin{align}
    J = \frac{eQ}{r^2}\delta(v) \ud v \;,
\end{align}
which is conserved
\be
\frac{1}{\sqrt{-g}}\partial_{\mu}(\sqrt{-g}J^{\mu})=0 \;,
\ee
as would also be the case for any field obtained by replacing $\delta(v)$  with some $F(v)$.

\paragraph{Appendix B:}\label{app:B}
The exact Bogoliubov coefficients computed from our wavefunction in \sv\, are $\alpha(\mathcal{E}',\mathcal{E})=\mathcal{F}(\mathcal{E}',\mathcal{E})$ and $\beta(\mathcal{E}',\mathcal{E})=\mathcal{F}(\mathcal{E}',-\mathcal{E})$, where
\begin{align}\label{eq:expact_bogol}
\mathcal{F}(\mathcal{E}', \mathcal{E}) = -8 \pi i e^{-i\Phi}e^{\pi g^2qQ} \, \frac{\Gamma(a) \Gamma(b)}{\Gamma(c)} \, {}_2 F_1\left(a,b;c; \frac{1}{\zeta - i 0^+ }\right) \, \left(\zeta - i 0^+ \right)^{-a}\,,
\end{align}
in which
\be
    \Phi=2 g^2 qQ \ln (2 \mathcal{E}' r_0) \;,
    \quad
    a = 1/2 + \mu - \kappa \,,
    \quad
    b = 3/2 + \mu - \kappa \;,
    \quad
    c = 1 + 2 \mu \;,
    \quad
    \zeta=\mathcal{E}/\mathcal{E}' \;.
\ee
As in the text, $\mu = i\sqrt{g^4 q^2 Q^2 - 1/4}$ and $\kappa = - i g^2 qQ$.

\begin{figure}[t]
\includegraphics[width=0.6\columnwidth]{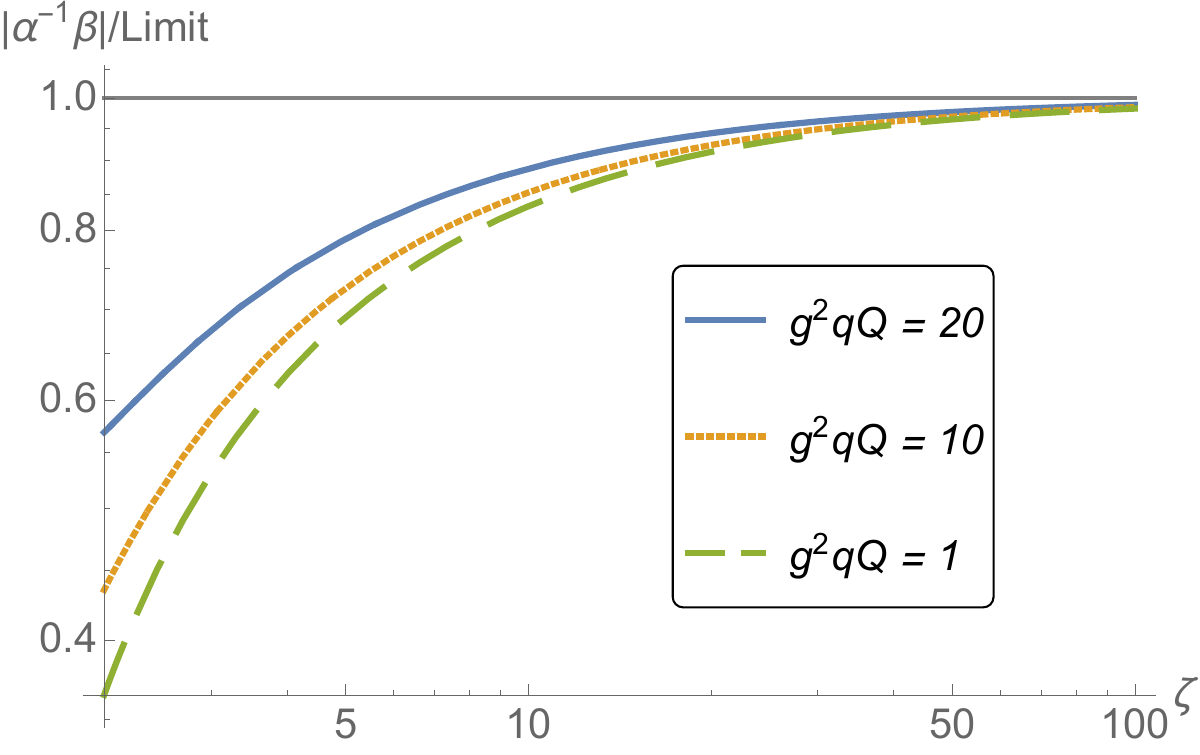}
\caption{\label{fig:amp} The exact result for the modulus of the pair creation amplitude $|\alpha^{-1}\beta|$ calculated from (\ref{eq:expact_bogol}), in ratio to its asymptotic limit (\ref{amp-gauge}), as a function of the ratio of energies $\zeta$, for three values of charge $g^2qQ$. The asymptotic limit is reached more quickly for larger values of the charge.}
\end{figure}

In the semiclassical limit $g^2 qQ \gg 1$, we can replace $\mu \to i g^2 qQ$. $\mathcal{F}$ also simplifies in the $|\mathcal{E}|\gg\mathcal{E}'$ limit, where the hypergeometric function evaluates to 1 and the otherwise complicated dependence of $\mathcal{F}(\mathcal{E}', \mathcal{E})$ on the ratio of energies, $\zeta$, reduces to a branch-cut structure captured by the final factor in (\ref{eq:expact_bogol}). It is this which gives rise to the exponential factor of $e^{-2\pi g^2qQ}$ between $|\alpha|$ and $|\beta|$ in (\ref{amp-gauge}). See Fig.~\ref{fig:amp}. 

Another interesting feature of \eqref{eq:expact_bogol} is that it also correctly captures the singularity at $\mathcal{E}\approx \mathcal{E}'$. This follows from the branch-cut of the hypergeometric function near $\zeta\rightarrow 1$ and leads to
\begin{align}
 \mathcal{F}(\mathcal{E}', \mathcal{E})&\sim \left(1-\zeta+i0^+\right)^{-1-2ig^2qQ}\quad;\quad \zeta\rightarrow 1\,,
\end{align} 
The semiclassical approximation, in fact, gives precisely the above branch-cut contribution, which is consistent with the observation of \cite{Aoude:2024sve} that the semiclassical wavefunction follows from the resummation of diagrams in the small-momentum-transfer limit.

\begin{thebibliography}{79}%
\makeatletter
\providecommand \@ifxundefined [1]{%
 \@ifx{#1\undefined}
}%
\providecommand \@ifnum [1]{%
 \ifnum #1\expandafter \@firstoftwo
 \else \expandafter \@secondoftwo
 \fi
}%
\providecommand \@ifx [1]{%
 \ifx #1\expandafter \@firstoftwo
 \else \expandafter \@secondoftwo
 \fi
}%
\providecommand \natexlab [1]{#1}%
\providecommand \enquote  [1]{``#1''}%
\providecommand \bibnamefont  [1]{#1}%
\providecommand \bibfnamefont [1]{#1}%
\providecommand \citenamefont [1]{#1}%
\providecommand \href@noop [0]{\@secondoftwo}%
\providecommand \href [0]{\begingroup \@sanitize@url \@href}%
\providecommand \@href[1]{\@@startlink{#1}\@@href}%
\providecommand \@@href[1]{\endgroup#1\@@endlink}%
\providecommand \@sanitize@url [0]{\catcode `\\12\catcode `\$12\catcode
  `\&12\catcode `\#12\catcode `\^12\catcode `\_12\catcode `\%12\relax}%
\providecommand \@@startlink[1]{}%
\providecommand \@@endlink[0]{}%
\providecommand \url  [0]{\begingroup\@sanitize@url \@url }%
\providecommand \@url [1]{\endgroup\@href {#1}{\urlprefix }}%
\providecommand \urlprefix  [0]{URL }%
\providecommand \Eprint [0]{\href }%
\providecommand \doibase [0]{http://dx.doi.org/}%
\providecommand \selectlanguage [0]{\@gobble}%
\providecommand \bibinfo  [0]{\@secondoftwo}%
\providecommand \bibfield  [0]{\@secondoftwo}%
\providecommand \translation [1]{[#1]}%
\providecommand \BibitemOpen [0]{}%
\providecommand \bibitemStop [0]{}%
\providecommand \bibitemNoStop [0]{.\EOS\space}%
\providecommand \EOS [0]{\spacefactor3000\relax}%
\providecommand \BibitemShut  [1]{\csname bibitem#1\endcsname}%
\let\auto@bib@innerbib\@empty
\bibitem [{\citenamefont {Bern}\ \emph {et~al.}(2024)\citenamefont {Bern},
  \citenamefont {Carrasco}, \citenamefont {Chiodaroli}, \citenamefont
  {Johansson},\ and\ \citenamefont {Roiban}}]{Bern:2019prr}%
  \BibitemOpen
  \bibfield  {author} {\bibinfo {author} {\bibfnamefont {Z.}~\bibnamefont
  {Bern}}, \bibinfo {author} {\bibfnamefont {J.~J.}\ \bibnamefont {Carrasco}},
  \bibinfo {author} {\bibfnamefont {M.}~\bibnamefont {Chiodaroli}}, \bibinfo
  {author} {\bibfnamefont {H.}~\bibnamefont {Johansson}}, \ and\ \bibinfo
  {author} {\bibfnamefont {R.}~\bibnamefont {Roiban}},\ }\href {\doibase
  10.1088/1751-8121/ad5fd0} {\bibfield  {journal} {\bibinfo  {journal} {J.
  Phys. A}\ }\textbf {\bibinfo {volume} {57}},\ \bibinfo {pages} {333002}
  (\bibinfo {year} {2024})},\ \Eprint {http://arxiv.org/abs/1909.01358}
  {arXiv:1909.01358 [hep-th]} \BibitemShut {NoStop}%
\bibitem [{\citenamefont {Borsten}(2020)}]{Borsten:2020bgv}%
  \BibitemOpen
  \bibfield  {author} {\bibinfo {author} {\bibfnamefont {L.}~\bibnamefont
  {Borsten}},\ }\href {\doibase 10.1007/s40766-020-00003-6} {\bibfield
  {journal} {\bibinfo  {journal} {Riv. Nuovo Cim.}\ }\textbf {\bibinfo {volume}
  {43}},\ \bibinfo {pages} {97} (\bibinfo {year} {2020})}\BibitemShut {NoStop}%
\bibitem [{\citenamefont {Bern}\ \emph {et~al.}(2022)\citenamefont {Bern},
  \citenamefont {Carrasco}, \citenamefont {Chiodaroli}, \citenamefont
  {Johansson},\ and\ \citenamefont {Roiban}}]{Bern:2022wqg}%
  \BibitemOpen
  \bibfield  {author} {\bibinfo {author} {\bibfnamefont {Z.}~\bibnamefont
  {Bern}}, \bibinfo {author} {\bibfnamefont {J.~J.}\ \bibnamefont {Carrasco}},
  \bibinfo {author} {\bibfnamefont {M.}~\bibnamefont {Chiodaroli}}, \bibinfo
  {author} {\bibfnamefont {H.}~\bibnamefont {Johansson}}, \ and\ \bibinfo
  {author} {\bibfnamefont {R.}~\bibnamefont {Roiban}},\ }\href {\doibase
  10.1088/1751-8121/ac93cf} {\bibfield  {journal} {\bibinfo  {journal} {J.
  Phys. A}\ }\textbf {\bibinfo {volume} {55}},\ \bibinfo {pages} {443003}
  (\bibinfo {year} {2022})},\ \Eprint {http://arxiv.org/abs/2203.13013}
  {arXiv:2203.13013 [hep-th]} \BibitemShut {NoStop}%
\bibitem [{\citenamefont {Adamo}\ \emph
  {et~al.}(2022{\natexlab{a}})\citenamefont {Adamo}, \citenamefont {Carrasco},
  \citenamefont {Carrillo-Gonz{\'a}lez}, \citenamefont {Chiodaroli},
  \citenamefont {Elvang}, \citenamefont {Johansson}, \citenamefont {O'Connell},
  \citenamefont {Roiban},\ and\ \citenamefont {Schlotterer}}]{Adamo:2022dcm}%
  \BibitemOpen
  \bibfield  {author} {\bibinfo {author} {\bibfnamefont {T.}~\bibnamefont
  {Adamo}}, \bibinfo {author} {\bibfnamefont {J.~J.~M.}\ \bibnamefont
  {Carrasco}}, \bibinfo {author} {\bibfnamefont {M.}~\bibnamefont
  {Carrillo-Gonz{\'a}lez}}, \bibinfo {author} {\bibfnamefont {M.}~\bibnamefont
  {Chiodaroli}}, \bibinfo {author} {\bibfnamefont {H.}~\bibnamefont {Elvang}},
  \bibinfo {author} {\bibfnamefont {H.}~\bibnamefont {Johansson}}, \bibinfo
  {author} {\bibfnamefont {D.}~\bibnamefont {O'Connell}}, \bibinfo {author}
  {\bibfnamefont {R.}~\bibnamefont {Roiban}}, \ and\ \bibinfo {author}
  {\bibfnamefont {O.}~\bibnamefont {Schlotterer}},\ }in\ \href@noop {} {\emph
  {\bibinfo {booktitle} {{Snowmass 2021}}}}\ (\bibinfo {year} {2022})\ \Eprint
  {http://arxiv.org/abs/2204.06547} {arXiv:2204.06547 [hep-th]} \BibitemShut
  {NoStop}%
\bibitem [{\citenamefont {Monteiro}\ and\ \citenamefont
  {O'Connell}(2011)}]{Monteiro:2011pc}%
  \BibitemOpen
  \bibfield  {author} {\bibinfo {author} {\bibfnamefont {R.}~\bibnamefont
  {Monteiro}}\ and\ \bibinfo {author} {\bibfnamefont {D.}~\bibnamefont
  {O'Connell}},\ }\href {\doibase 10.1007/JHEP07(2011)007} {\bibfield
  {journal} {\bibinfo  {journal} {JHEP}\ }\textbf {\bibinfo {volume} {07}},\
  \bibinfo {pages} {007} (\bibinfo {year} {2011})},\ \Eprint
  {http://arxiv.org/abs/1105.2565} {arXiv:1105.2565 [hep-th]} \BibitemShut
  {NoStop}%
\bibitem [{\citenamefont {Huang}\ \emph {et~al.}(2020)\citenamefont {Huang},
  \citenamefont {Kol},\ and\ \citenamefont {O'Connell}}]{Huang:2019cja}%
  \BibitemOpen
  \bibfield  {author} {\bibinfo {author} {\bibfnamefont {Y.-T.}\ \bibnamefont
  {Huang}}, \bibinfo {author} {\bibfnamefont {U.}~\bibnamefont {Kol}}, \ and\
  \bibinfo {author} {\bibfnamefont {D.}~\bibnamefont {O'Connell}},\ }\href
  {\doibase 10.1103/PhysRevD.102.046005} {\bibfield  {journal} {\bibinfo
  {journal} {Phys. Rev. D}\ }\textbf {\bibinfo {volume} {102}},\ \bibinfo
  {pages} {046005} (\bibinfo {year} {2020})},\ \Eprint
  {http://arxiv.org/abs/1911.06318} {arXiv:1911.06318 [hep-th]} \BibitemShut
  {NoStop}%
\bibitem [{\citenamefont {Alawadhi}\ \emph {et~al.}(2020)\citenamefont
  {Alawadhi}, \citenamefont {Berman}, \citenamefont {Spence},\ and\
  \citenamefont {Peinador~Veiga}}]{Alawadhi:2019urr}%
  \BibitemOpen
  \bibfield  {author} {\bibinfo {author} {\bibfnamefont {R.}~\bibnamefont
  {Alawadhi}}, \bibinfo {author} {\bibfnamefont {D.~S.}\ \bibnamefont
  {Berman}}, \bibinfo {author} {\bibfnamefont {B.}~\bibnamefont {Spence}}, \
  and\ \bibinfo {author} {\bibfnamefont {D.}~\bibnamefont {Peinador~Veiga}},\
  }\href {\doibase 10.1007/JHEP03(2020)059} {\bibfield  {journal} {\bibinfo
  {journal} {JHEP}\ }\textbf {\bibinfo {volume} {03}},\ \bibinfo {pages} {059}
  (\bibinfo {year} {2020})},\ \Eprint {http://arxiv.org/abs/1911.06797}
  {arXiv:1911.06797 [hep-th]} \BibitemShut {NoStop}%
\bibitem [{\citenamefont {Banerjee}\ \emph {et~al.}(2020)\citenamefont
  {Banerjee}, \citenamefont {Colg{\'a}in}, \citenamefont {Rosabal},\ and\
  \citenamefont {Yavartanoo}}]{Banerjee:2019saj}%
  \BibitemOpen
  \bibfield  {author} {\bibinfo {author} {\bibfnamefont {A.}~\bibnamefont
  {Banerjee}}, \bibinfo {author} {\bibfnamefont {E.~{\'O}.}\ \bibnamefont
  {Colg{\'a}in}}, \bibinfo {author} {\bibfnamefont {J.~A.}\ \bibnamefont
  {Rosabal}}, \ and\ \bibinfo {author} {\bibfnamefont {H.}~\bibnamefont
  {Yavartanoo}},\ }\href {\doibase 10.1103/PhysRevD.102.126017} {\bibfield
  {journal} {\bibinfo  {journal} {Phys. Rev. D}\ }\textbf {\bibinfo {volume}
  {102}},\ \bibinfo {pages} {126017} (\bibinfo {year} {2020})},\ \Eprint
  {http://arxiv.org/abs/1912.02597} {arXiv:1912.02597 [hep-th]} \BibitemShut
  {NoStop}%
\bibitem [{\citenamefont {Cheung}\ \emph
  {et~al.}(2022{\natexlab{a}})\citenamefont {Cheung}, \citenamefont {Mangan},
  \citenamefont {Parra-Martinez},\ and\ \citenamefont {Shah}}]{Cheung:2022mix}%
  \BibitemOpen
  \bibfield  {author} {\bibinfo {author} {\bibfnamefont {C.}~\bibnamefont
  {Cheung}}, \bibinfo {author} {\bibfnamefont {J.}~\bibnamefont {Mangan}},
  \bibinfo {author} {\bibfnamefont {J.}~\bibnamefont {Parra-Martinez}}, \ and\
  \bibinfo {author} {\bibfnamefont {N.}~\bibnamefont {Shah}},\ }\href {\doibase
  10.1103/PhysRevLett.129.221602} {\bibfield  {journal} {\bibinfo  {journal}
  {Phys. Rev. Lett.}\ }\textbf {\bibinfo {volume} {129}},\ \bibinfo {pages}
  {221602} (\bibinfo {year} {2022}{\natexlab{a}})},\ \Eprint
  {http://arxiv.org/abs/2204.07130} {arXiv:2204.07130 [hep-th]} \BibitemShut
  {NoStop}%
\bibitem [{\citenamefont {Borsten}\ \emph {et~al.}(2021)\citenamefont
  {Borsten}, \citenamefont {Kim}, \citenamefont {Jur{\v{c}}o}, \citenamefont
  {Macrelli}, \citenamefont {Saemann},\ and\ \citenamefont
  {Wolf}}]{Borsten:2021hua}%
  \BibitemOpen
  \bibfield  {author} {\bibinfo {author} {\bibfnamefont {L.}~\bibnamefont
  {Borsten}}, \bibinfo {author} {\bibfnamefont {H.}~\bibnamefont {Kim}},
  \bibinfo {author} {\bibfnamefont {B.}~\bibnamefont {Jur{\v{c}}o}}, \bibinfo
  {author} {\bibfnamefont {T.}~\bibnamefont {Macrelli}}, \bibinfo {author}
  {\bibfnamefont {C.}~\bibnamefont {Saemann}}, \ and\ \bibinfo {author}
  {\bibfnamefont {M.}~\bibnamefont {Wolf}},\ }\href {\doibase
  10.1002/prop.202100075} {\bibfield  {journal} {\bibinfo  {journal} {Fortsch.
  Phys.}\ }\textbf {\bibinfo {volume} {69}},\ \bibinfo {pages} {2100075}
  (\bibinfo {year} {2021})},\ \Eprint {http://arxiv.org/abs/2102.11390}
  {arXiv:2102.11390 [hep-th]} \BibitemShut {NoStop}%
\bibitem [{\citenamefont {Alawadhi}\ \emph {et~al.}(2021)\citenamefont
  {Alawadhi}, \citenamefont {Berman}, \citenamefont {White},\ and\
  \citenamefont {Wikeley}}]{Alawadhi:2021uie}%
  \BibitemOpen
  \bibfield  {author} {\bibinfo {author} {\bibfnamefont {R.}~\bibnamefont
  {Alawadhi}}, \bibinfo {author} {\bibfnamefont {D.~S.}\ \bibnamefont
  {Berman}}, \bibinfo {author} {\bibfnamefont {C.~D.}\ \bibnamefont {White}}, \
  and\ \bibinfo {author} {\bibfnamefont {S.}~\bibnamefont {Wikeley}},\ }\href
  {\doibase 10.1007/JHEP10(2021)229} {\bibfield  {journal} {\bibinfo  {journal}
  {JHEP}\ }\textbf {\bibinfo {volume} {10}},\ \bibinfo {pages} {229} (\bibinfo
  {year} {2021})},\ \Eprint {http://arxiv.org/abs/2107.01114} {arXiv:2107.01114
  [hep-th]} \BibitemShut {NoStop}%
\bibitem [{\citenamefont {Armstrong-Williams}\ \emph
  {et~al.}(2022)\citenamefont {Armstrong-Williams}, \citenamefont {White},\
  and\ \citenamefont {Wikeley}}]{Armstrong-Williams:2022apo}%
  \BibitemOpen
  \bibfield  {author} {\bibinfo {author} {\bibfnamefont {K.}~\bibnamefont
  {Armstrong-Williams}}, \bibinfo {author} {\bibfnamefont {C.~D.}\ \bibnamefont
  {White}}, \ and\ \bibinfo {author} {\bibfnamefont {S.}~\bibnamefont
  {Wikeley}},\ }\href {\doibase 10.1007/JHEP08(2022)160} {\bibfield  {journal}
  {\bibinfo  {journal} {JHEP}\ }\textbf {\bibinfo {volume} {08}},\ \bibinfo
  {pages} {160} (\bibinfo {year} {2022})},\ \Eprint
  {http://arxiv.org/abs/2205.02136} {arXiv:2205.02136 [hep-th]} \BibitemShut
  {NoStop}%
\bibitem [{\citenamefont {Luna}\ \emph {et~al.}(2017)\citenamefont {Luna},
  \citenamefont {Monteiro}, \citenamefont {Nicholson}, \citenamefont {Ochirov},
  \citenamefont {O'Connell}, \citenamefont {Westerberg},\ and\ \citenamefont
  {White}}]{Luna:2016hge}%
  \BibitemOpen
  \bibfield  {author} {\bibinfo {author} {\bibfnamefont {A.}~\bibnamefont
  {Luna}}, \bibinfo {author} {\bibfnamefont {R.}~\bibnamefont {Monteiro}},
  \bibinfo {author} {\bibfnamefont {I.}~\bibnamefont {Nicholson}}, \bibinfo
  {author} {\bibfnamefont {A.}~\bibnamefont {Ochirov}}, \bibinfo {author}
  {\bibfnamefont {D.}~\bibnamefont {O'Connell}}, \bibinfo {author}
  {\bibfnamefont {N.}~\bibnamefont {Westerberg}}, \ and\ \bibinfo {author}
  {\bibfnamefont {C.~D.}\ \bibnamefont {White}},\ }\href {\doibase
  10.1007/JHEP04(2017)069} {\bibfield  {journal} {\bibinfo  {journal} {JHEP}\
  }\textbf {\bibinfo {volume} {04}},\ \bibinfo {pages} {069} (\bibinfo {year}
  {2017})},\ \Eprint {http://arxiv.org/abs/1611.07508} {arXiv:1611.07508
  [hep-th]} \BibitemShut {NoStop}%
\bibitem [{\citenamefont {Adamo}\ \emph {et~al.}(2018)\citenamefont {Adamo},
  \citenamefont {Casali}, \citenamefont {Mason},\ and\ \citenamefont
  {Nekovar}}]{Adamo:2017nia}%
  \BibitemOpen
  \bibfield  {author} {\bibinfo {author} {\bibfnamefont {T.}~\bibnamefont
  {Adamo}}, \bibinfo {author} {\bibfnamefont {E.}~\bibnamefont {Casali}},
  \bibinfo {author} {\bibfnamefont {L.}~\bibnamefont {Mason}}, \ and\ \bibinfo
  {author} {\bibfnamefont {S.}~\bibnamefont {Nekovar}},\ }\href {\doibase
  10.1088/1361-6382/aa9961} {\bibfield  {journal} {\bibinfo  {journal} {Class.
  Quant. Grav.}\ }\textbf {\bibinfo {volume} {35}},\ \bibinfo {pages} {015004}
  (\bibinfo {year} {2018})},\ \Eprint {http://arxiv.org/abs/1706.08925}
  {arXiv:1706.08925 [hep-th]} \BibitemShut {NoStop}%
\bibitem [{\citenamefont {Albayrak}\ \emph {et~al.}(2021)\citenamefont
  {Albayrak}, \citenamefont {Kharel},\ and\ \citenamefont
  {Meltzer}}]{Albayrak:2020fyp}%
  \BibitemOpen
  \bibfield  {author} {\bibinfo {author} {\bibfnamefont {S.}~\bibnamefont
  {Albayrak}}, \bibinfo {author} {\bibfnamefont {S.}~\bibnamefont {Kharel}}, \
  and\ \bibinfo {author} {\bibfnamefont {D.}~\bibnamefont {Meltzer}},\ }\href
  {\doibase 10.1007/JHEP03(2021)249} {\bibfield  {journal} {\bibinfo  {journal}
  {JHEP}\ }\textbf {\bibinfo {volume} {03}},\ \bibinfo {pages} {249} (\bibinfo
  {year} {2021})},\ \Eprint {http://arxiv.org/abs/2012.10460} {arXiv:2012.10460
  [hep-th]} \BibitemShut {NoStop}%
\bibitem [{\citenamefont {Adamo}\ and\ \citenamefont
  {Ilderton}(2020)}]{Adamo:2020qru}%
  \BibitemOpen
  \bibfield  {author} {\bibinfo {author} {\bibfnamefont {T.}~\bibnamefont
  {Adamo}}\ and\ \bibinfo {author} {\bibfnamefont {A.}~\bibnamefont
  {Ilderton}},\ }\href {\doibase 10.1007/JHEP09(2020)200} {\bibfield  {journal}
  {\bibinfo  {journal} {JHEP}\ }\textbf {\bibinfo {volume} {09}},\ \bibinfo
  {pages} {200} (\bibinfo {year} {2020})},\ \Eprint
  {http://arxiv.org/abs/2005.05807} {arXiv:2005.05807 [hep-th]} \BibitemShut
  {NoStop}%
\bibitem [{\citenamefont {Zhou}(2021)}]{Zhou:2021gnu}%
  \BibitemOpen
  \bibfield  {author} {\bibinfo {author} {\bibfnamefont {X.}~\bibnamefont
  {Zhou}},\ }\href {\doibase 10.1103/PhysRevLett.127.141601} {\bibfield
  {journal} {\bibinfo  {journal} {Phys. Rev. Lett.}\ }\textbf {\bibinfo
  {volume} {127}},\ \bibinfo {pages} {141601} (\bibinfo {year} {2021})},\
  \Eprint {http://arxiv.org/abs/2106.07651} {arXiv:2106.07651 [hep-th]}
  \BibitemShut {NoStop}%
\bibitem [{\citenamefont {Diwakar}\ \emph {et~al.}(2021)\citenamefont
  {Diwakar}, \citenamefont {Herderschee}, \citenamefont {Roiban},\ and\
  \citenamefont {Teng}}]{Diwakar:2021juk}%
  \BibitemOpen
  \bibfield  {author} {\bibinfo {author} {\bibfnamefont {P.}~\bibnamefont
  {Diwakar}}, \bibinfo {author} {\bibfnamefont {A.}~\bibnamefont
  {Herderschee}}, \bibinfo {author} {\bibfnamefont {R.}~\bibnamefont {Roiban}},
  \ and\ \bibinfo {author} {\bibfnamefont {F.}~\bibnamefont {Teng}},\ }\href
  {\doibase 10.1007/JHEP10(2021)141} {\bibfield  {journal} {\bibinfo  {journal}
  {JHEP}\ }\textbf {\bibinfo {volume} {10}},\ \bibinfo {pages} {141} (\bibinfo
  {year} {2021})},\ \Eprint {http://arxiv.org/abs/2106.10822} {arXiv:2106.10822
  [hep-th]} \BibitemShut {NoStop}%
\bibitem [{\citenamefont {Cheung}\ \emph
  {et~al.}(2022{\natexlab{b}})\citenamefont {Cheung}, \citenamefont
  {Parra-Martinez},\ and\ \citenamefont {Sivaramakrishnan}}]{Cheung:2022pdk}%
  \BibitemOpen
  \bibfield  {author} {\bibinfo {author} {\bibfnamefont {C.}~\bibnamefont
  {Cheung}}, \bibinfo {author} {\bibfnamefont {J.}~\bibnamefont
  {Parra-Martinez}}, \ and\ \bibinfo {author} {\bibfnamefont {A.}~\bibnamefont
  {Sivaramakrishnan}},\ }\href {\doibase 10.1007/JHEP05(2022)027} {\bibfield
  {journal} {\bibinfo  {journal} {JHEP}\ }\textbf {\bibinfo {volume} {05}},\
  \bibinfo {pages} {027} (\bibinfo {year} {2022}{\natexlab{b}})},\ \Eprint
  {http://arxiv.org/abs/2201.05147} {arXiv:2201.05147 [hep-th]} \BibitemShut
  {NoStop}%
\bibitem [{\citenamefont {Herderschee}\ \emph {et~al.}(2022)\citenamefont
  {Herderschee}, \citenamefont {Roiban},\ and\ \citenamefont
  {Teng}}]{Herderschee:2022ntr}%
  \BibitemOpen
  \bibfield  {author} {\bibinfo {author} {\bibfnamefont {A.}~\bibnamefont
  {Herderschee}}, \bibinfo {author} {\bibfnamefont {R.}~\bibnamefont {Roiban}},
  \ and\ \bibinfo {author} {\bibfnamefont {F.}~\bibnamefont {Teng}},\ }\href
  {\doibase 10.1007/JHEP05(2022)026} {\bibfield  {journal} {\bibinfo  {journal}
  {JHEP}\ }\textbf {\bibinfo {volume} {05}},\ \bibinfo {pages} {026} (\bibinfo
  {year} {2022})},\ \Eprint {http://arxiv.org/abs/2201.05067} {arXiv:2201.05067
  [hep-th]} \BibitemShut {NoStop}%
\bibitem [{\citenamefont {Drummond}\ \emph {et~al.}(2023)\citenamefont
  {Drummond}, \citenamefont {Glew},\ and\ \citenamefont
  {Santagata}}]{Drummond:2022dxd}%
  \BibitemOpen
  \bibfield  {author} {\bibinfo {author} {\bibfnamefont {J.~M.}\ \bibnamefont
  {Drummond}}, \bibinfo {author} {\bibfnamefont {R.}~\bibnamefont {Glew}}, \
  and\ \bibinfo {author} {\bibfnamefont {M.}~\bibnamefont {Santagata}},\ }\href
  {\doibase 10.1103/PhysRevD.107.L081901} {\bibfield  {journal} {\bibinfo
  {journal} {Phys. Rev. D}\ }\textbf {\bibinfo {volume} {107}},\ \bibinfo
  {pages} {L081901} (\bibinfo {year} {2023})},\ \Eprint
  {http://arxiv.org/abs/2202.09837} {arXiv:2202.09837 [hep-th]} \BibitemShut
  {NoStop}%
\bibitem [{\citenamefont {Lee}\ and\ \citenamefont {Wang}(2023)}]{Lee:2022fgr}%
  \BibitemOpen
  \bibfield  {author} {\bibinfo {author} {\bibfnamefont {H.}~\bibnamefont
  {Lee}}\ and\ \bibinfo {author} {\bibfnamefont {X.}~\bibnamefont {Wang}},\
  }\href {\doibase 10.1103/PhysRevD.108.L061702} {\bibfield  {journal}
  {\bibinfo  {journal} {Phys. Rev. D}\ }\textbf {\bibinfo {volume} {108}},\
  \bibinfo {pages} {L061702} (\bibinfo {year} {2023})},\ \Eprint
  {http://arxiv.org/abs/2212.11282} {arXiv:2212.11282 [hep-th]} \BibitemShut
  {NoStop}%
\bibitem [{\citenamefont {Lipstein}\ and\ \citenamefont
  {Nagy}(2023)}]{Lipstein:2023pih}%
  \BibitemOpen
  \bibfield  {author} {\bibinfo {author} {\bibfnamefont {A.}~\bibnamefont
  {Lipstein}}\ and\ \bibinfo {author} {\bibfnamefont {S.}~\bibnamefont
  {Nagy}},\ }\href {\doibase 10.1103/PhysRevLett.131.081501} {\bibfield
  {journal} {\bibinfo  {journal} {Phys. Rev. Lett.}\ }\textbf {\bibinfo
  {volume} {131}},\ \bibinfo {pages} {081501} (\bibinfo {year} {2023})},\
  \Eprint {http://arxiv.org/abs/2304.07141} {arXiv:2304.07141 [hep-th]}
  \BibitemShut {NoStop}%
\bibitem [{\citenamefont {Mei}(2023)}]{Mei:2023jkb}%
  \BibitemOpen
  \bibfield  {author} {\bibinfo {author} {\bibfnamefont {J.}~\bibnamefont
  {Mei}},\ }\href@noop {} {\  (\bibinfo {year} {2023})},\ \Eprint
  {http://arxiv.org/abs/2305.13894} {arXiv:2305.13894 [hep-th]} \BibitemShut
  {NoStop}%
\bibitem [{\citenamefont {Liang}\ and\ \citenamefont
  {Nagy}(2024)}]{Liang:2023zxo}%
  \BibitemOpen
  \bibfield  {author} {\bibinfo {author} {\bibfnamefont {Q.}~\bibnamefont
  {Liang}}\ and\ \bibinfo {author} {\bibfnamefont {S.}~\bibnamefont {Nagy}},\
  }\href {\doibase 10.1007/JHEP04(2024)139} {\bibfield  {journal} {\bibinfo
  {journal} {JHEP}\ }\textbf {\bibinfo {volume} {04}},\ \bibinfo {pages} {139}
  (\bibinfo {year} {2024})},\ \Eprint {http://arxiv.org/abs/2311.14319}
  {arXiv:2311.14319 [hep-th]} \BibitemShut {NoStop}%
\bibitem [{\citenamefont {Brown}\ \emph {et~al.}(2024)\citenamefont {Brown},
  \citenamefont {Gowdy},\ and\ \citenamefont {Spence}}]{Brown:2023zxm}%
  \BibitemOpen
  \bibfield  {author} {\bibinfo {author} {\bibfnamefont {G.~R.}\ \bibnamefont
  {Brown}}, \bibinfo {author} {\bibfnamefont {J.}~\bibnamefont {Gowdy}}, \ and\
  \bibinfo {author} {\bibfnamefont {B.}~\bibnamefont {Spence}},\ }\href
  {\doibase 10.1103/PhysRevD.109.026009} {\bibfield  {journal} {\bibinfo
  {journal} {Phys. Rev. D}\ }\textbf {\bibinfo {volume} {109}},\ \bibinfo
  {pages} {026009} (\bibinfo {year} {2024})},\ \Eprint
  {http://arxiv.org/abs/2307.11063} {arXiv:2307.11063 [hep-th]} \BibitemShut
  {NoStop}%
\bibitem [{\citenamefont {Carrillo~Gonz{\'a}lez}\ \emph
  {et~al.}(2024)\citenamefont {Carrillo~Gonz{\'a}lez}, \citenamefont
  {Lipstein},\ and\ \citenamefont {Nagy}}]{CarrilloGonzalez:2024sto}%
  \BibitemOpen
  \bibfield  {author} {\bibinfo {author} {\bibfnamefont {M.}~\bibnamefont
  {Carrillo~Gonz{\'a}lez}}, \bibinfo {author} {\bibfnamefont {A.}~\bibnamefont
  {Lipstein}}, \ and\ \bibinfo {author} {\bibfnamefont {S.}~\bibnamefont
  {Nagy}},\ }\href {\doibase 10.1007/JHEP10(2024)183} {\bibfield  {journal}
  {\bibinfo  {journal} {JHEP}\ }\textbf {\bibinfo {volume} {10}},\ \bibinfo
  {pages} {183} (\bibinfo {year} {2024})},\ \Eprint
  {http://arxiv.org/abs/2407.12905} {arXiv:2407.12905 [hep-th]} \BibitemShut
  {NoStop}%
\bibitem [{\citenamefont {Beetar}\ \emph {et~al.}(2025)\citenamefont {Beetar},
  \citenamefont {Carrillo~Gonz{\'a}lez}, \citenamefont {Jaitly},\ and\
  \citenamefont {Keseman}}]{Beetar:2024ptv}%
  \BibitemOpen
  \bibfield  {author} {\bibinfo {author} {\bibfnamefont {C.}~\bibnamefont
  {Beetar}}, \bibinfo {author} {\bibfnamefont {M.}~\bibnamefont
  {Carrillo~Gonz{\'a}lez}}, \bibinfo {author} {\bibfnamefont {S.}~\bibnamefont
  {Jaitly}}, \ and\ \bibinfo {author} {\bibfnamefont {T.}~\bibnamefont
  {Keseman}},\ }\href {\doibase 10.1007/JHEP03(2025)125} {\bibfield  {journal}
  {\bibinfo  {journal} {JHEP}\ }\textbf {\bibinfo {volume} {03}},\ \bibinfo
  {pages} {125} (\bibinfo {year} {2025})},\ \Eprint
  {http://arxiv.org/abs/2410.23342} {arXiv:2410.23342 [hep-th]} \BibitemShut
  {NoStop}%
\bibitem [{\citenamefont {Ilderton}\ and\ \citenamefont
  {Lindved}(2024)}]{Ilderton:2024oly}%
  \BibitemOpen
  \bibfield  {author} {\bibinfo {author} {\bibfnamefont {A.}~\bibnamefont
  {Ilderton}}\ and\ \bibinfo {author} {\bibfnamefont {W.}~\bibnamefont
  {Lindved}},\ }\href {\doibase 10.1007/JHEP11(2024)100} {\bibfield  {journal}
  {\bibinfo  {journal} {JHEP}\ }\textbf {\bibinfo {volume} {11}},\ \bibinfo
  {pages} {100} (\bibinfo {year} {2024})},\ \Eprint
  {http://arxiv.org/abs/2405.10016} {arXiv:2405.10016 [hep-th]} \BibitemShut
  {NoStop}%
\bibitem [{\citenamefont {Alday}\ \emph {et~al.}(2025)\citenamefont {Alday},
  \citenamefont {Nocchi},\ and\ \citenamefont {Sangar{\'e}}}]{Alday:2025bjp}%
  \BibitemOpen
  \bibfield  {author} {\bibinfo {author} {\bibfnamefont {L.~F.}\ \bibnamefont
  {Alday}}, \bibinfo {author} {\bibfnamefont {M.}~\bibnamefont {Nocchi}}, \
  and\ \bibinfo {author} {\bibfnamefont {A.~S.}\ \bibnamefont {Sangar{\'e}}},\
  }\href@noop {} {\  (\bibinfo {year} {2025})},\ \Eprint
  {http://arxiv.org/abs/2504.19973} {arXiv:2504.19973 [hep-th]} \BibitemShut
  {NoStop}%
\bibitem [{\citenamefont {Ilderton}\ and\ \citenamefont
  {Lindved}(2025)}]{Ilderton:2025gug}%
  \BibitemOpen
  \bibfield  {author} {\bibinfo {author} {\bibfnamefont {A.}~\bibnamefont
  {Ilderton}}\ and\ \bibinfo {author} {\bibfnamefont {W.}~\bibnamefont
  {Lindved}},\ }\href {\doibase 10.1007/JHEP09(2025)156} {\bibfield  {journal}
  {\bibinfo  {journal} {JHEP}\ }\textbf {\bibinfo {volume} {09}},\ \bibinfo
  {pages} {156} (\bibinfo {year} {2025})},\ \Eprint
  {http://arxiv.org/abs/2505.16852} {arXiv:2505.16852 [hep-th]} \BibitemShut
  {NoStop}%
\bibitem [{\citenamefont {Monteiro}\ \emph {et~al.}(2014)\citenamefont
  {Monteiro}, \citenamefont {O'Connell},\ and\ \citenamefont
  {White}}]{Monteiro:2014cda}%
  \BibitemOpen
  \bibfield  {author} {\bibinfo {author} {\bibfnamefont {R.}~\bibnamefont
  {Monteiro}}, \bibinfo {author} {\bibfnamefont {D.}~\bibnamefont {O'Connell}},
  \ and\ \bibinfo {author} {\bibfnamefont {C.~D.}\ \bibnamefont {White}},\
  }\href {\doibase 10.1007/JHEP12(2014)056} {\bibfield  {journal} {\bibinfo
  {journal} {JHEP}\ }\textbf {\bibinfo {volume} {12}},\ \bibinfo {pages} {056}
  (\bibinfo {year} {2014})},\ \Eprint {http://arxiv.org/abs/1410.0239}
  {arXiv:1410.0239 [hep-th]} \BibitemShut {NoStop}%
\bibitem [{\citenamefont {Luna}\ \emph {et~al.}(2015)\citenamefont {Luna},
  \citenamefont {Monteiro}, \citenamefont {O'Connell},\ and\ \citenamefont
  {White}}]{Luna:2015paa}%
  \BibitemOpen
  \bibfield  {author} {\bibinfo {author} {\bibfnamefont {A.}~\bibnamefont
  {Luna}}, \bibinfo {author} {\bibfnamefont {R.}~\bibnamefont {Monteiro}},
  \bibinfo {author} {\bibfnamefont {D.}~\bibnamefont {O'Connell}}, \ and\
  \bibinfo {author} {\bibfnamefont {C.~D.}\ \bibnamefont {White}},\ }\href
  {\doibase 10.1016/j.physletb.2015.09.021} {\bibfield  {journal} {\bibinfo
  {journal} {Phys. Lett. B}\ }\textbf {\bibinfo {volume} {750}},\ \bibinfo
  {pages} {272} (\bibinfo {year} {2015})},\ \Eprint
  {http://arxiv.org/abs/1507.01869} {arXiv:1507.01869 [hep-th]} \BibitemShut
  {NoStop}%
\bibitem [{\citenamefont {Berman}\ \emph {et~al.}(2019)\citenamefont {Berman},
  \citenamefont {Chac{\'o}n}, \citenamefont {Luna},\ and\ \citenamefont
  {White}}]{Berman:2018hwd}%
  \BibitemOpen
  \bibfield  {author} {\bibinfo {author} {\bibfnamefont {D.~S.}\ \bibnamefont
  {Berman}}, \bibinfo {author} {\bibfnamefont {E.}~\bibnamefont {Chac{\'o}n}},
  \bibinfo {author} {\bibfnamefont {A.}~\bibnamefont {Luna}}, \ and\ \bibinfo
  {author} {\bibfnamefont {C.~D.}\ \bibnamefont {White}},\ }\href {\doibase
  10.1007/JHEP01(2019)107} {\bibfield  {journal} {\bibinfo  {journal} {JHEP}\
  }\textbf {\bibinfo {volume} {01}},\ \bibinfo {pages} {107} (\bibinfo {year}
  {2019})},\ \Eprint {http://arxiv.org/abs/1809.04063} {arXiv:1809.04063
  [hep-th]} \BibitemShut {NoStop}%
\bibitem [{\citenamefont {Alfonsi}\ \emph {et~al.}(2020)\citenamefont
  {Alfonsi}, \citenamefont {White},\ and\ \citenamefont
  {Wikeley}}]{Alfonsi:2020lub}%
  \BibitemOpen
  \bibfield  {author} {\bibinfo {author} {\bibfnamefont {L.}~\bibnamefont
  {Alfonsi}}, \bibinfo {author} {\bibfnamefont {C.~D.}\ \bibnamefont {White}},
  \ and\ \bibinfo {author} {\bibfnamefont {S.}~\bibnamefont {Wikeley}},\ }\href
  {\doibase 10.1007/JHEP07(2020)091} {\bibfield  {journal} {\bibinfo  {journal}
  {JHEP}\ }\textbf {\bibinfo {volume} {07}},\ \bibinfo {pages} {091} (\bibinfo
  {year} {2020})},\ \Eprint {http://arxiv.org/abs/2004.07181} {arXiv:2004.07181
  [hep-th]} \BibitemShut {NoStop}%
\bibitem [{\citenamefont {Bahjat-Abbas}\ \emph {et~al.}(2020)\citenamefont
  {Bahjat-Abbas}, \citenamefont {Stark-Much{\~a}o},\ and\ \citenamefont
  {White}}]{Bahjat-Abbas:2020cyb}%
  \BibitemOpen
  \bibfield  {author} {\bibinfo {author} {\bibfnamefont {N.}~\bibnamefont
  {Bahjat-Abbas}}, \bibinfo {author} {\bibfnamefont {R.}~\bibnamefont
  {Stark-Much{\~a}o}}, \ and\ \bibinfo {author} {\bibfnamefont {C.~D.}\
  \bibnamefont {White}},\ }\href {\doibase 10.1007/JHEP04(2020)102} {\bibfield
  {journal} {\bibinfo  {journal} {JHEP}\ }\textbf {\bibinfo {volume} {04}},\
  \bibinfo {pages} {102} (\bibinfo {year} {2020})},\ \Eprint
  {http://arxiv.org/abs/2001.09918} {arXiv:2001.09918 [hep-th]} \BibitemShut
  {NoStop}%
\bibitem [{\citenamefont {Adamo}\ and\ \citenamefont
  {Kol}(2022)}]{Adamo:2021dfg}%
  \BibitemOpen
  \bibfield  {author} {\bibinfo {author} {\bibfnamefont {T.}~\bibnamefont
  {Adamo}}\ and\ \bibinfo {author} {\bibfnamefont {U.}~\bibnamefont {Kol}},\
  }\href {\doibase 10.1088/1361-6382/ac635e} {\bibfield  {journal} {\bibinfo
  {journal} {Class. Quant. Grav.}\ }\textbf {\bibinfo {volume} {39}},\ \bibinfo
  {pages} {105007} (\bibinfo {year} {2022})},\ \Eprint
  {http://arxiv.org/abs/2109.07832} {arXiv:2109.07832 [hep-th]} \BibitemShut
  {NoStop}%
\bibitem [{\citenamefont {Luna}\ \emph {et~al.}(2022)\citenamefont {Luna},
  \citenamefont {Moynihan},\ and\ \citenamefont {White}}]{Luna:2022dxo}%
  \BibitemOpen
  \bibfield  {author} {\bibinfo {author} {\bibfnamefont {A.}~\bibnamefont
  {Luna}}, \bibinfo {author} {\bibfnamefont {N.}~\bibnamefont {Moynihan}}, \
  and\ \bibinfo {author} {\bibfnamefont {C.~D.}\ \bibnamefont {White}},\ }\href
  {\doibase 10.1007/JHEP12(2022)046} {\bibfield  {journal} {\bibinfo  {journal}
  {JHEP}\ }\textbf {\bibinfo {volume} {12}},\ \bibinfo {pages} {046} (\bibinfo
  {year} {2022})},\ \Eprint {http://arxiv.org/abs/2208.08548} {arXiv:2208.08548
  [hep-th]} \BibitemShut {NoStop}%
\bibitem [{\citenamefont {White}(2024)}]{White:2024pve}%
  \BibitemOpen
  \bibfield  {author} {\bibinfo {author} {\bibfnamefont {C.~D.}\ \bibnamefont
  {White}},\ }\href {\doibase 10.1142/q0457} {\emph {\bibinfo {title} {{The
  Classical Double Copy}}}}\ (\bibinfo  {publisher} {World Scientific},\
  \bibinfo {year} {2024})\BibitemShut {NoStop}%
\bibitem [{\citenamefont {Kim}(2025)}]{Kim:2024dxo}%
  \BibitemOpen
  \bibfield  {author} {\bibinfo {author} {\bibfnamefont {J.-H.}\ \bibnamefont
  {Kim}},\ }\href {\doibase 10.1103/PhysRevD.111.L021703} {\bibfield  {journal}
  {\bibinfo  {journal} {Phys. Rev. D}\ }\textbf {\bibinfo {volume} {111}},\
  \bibinfo {pages} {L021703} (\bibinfo {year} {2025})},\ \Eprint
  {http://arxiv.org/abs/2405.09518} {arXiv:2405.09518 [hep-th]} \BibitemShut
  {NoStop}%
\bibitem [{\citenamefont {Chawla}\ \emph {et~al.}(2024)\citenamefont {Chawla},
  \citenamefont {Fransen},\ and\ \citenamefont {Keeler}}]{Chawla:2024mse}%
  \BibitemOpen
  \bibfield  {author} {\bibinfo {author} {\bibfnamefont {S.}~\bibnamefont
  {Chawla}}, \bibinfo {author} {\bibfnamefont {K.}~\bibnamefont {Fransen}}, \
  and\ \bibinfo {author} {\bibfnamefont {C.}~\bibnamefont {Keeler}},\ }\href
  {\doibase 10.1088/1361-6382/ad8f8c} {\bibfield  {journal} {\bibinfo
  {journal} {Class. Quant. Grav.}\ }\textbf {\bibinfo {volume} {41}},\ \bibinfo
  {pages} {245015} (\bibinfo {year} {2024})},\ \Eprint
  {http://arxiv.org/abs/2406.14601} {arXiv:2406.14601 [hep-th]} \BibitemShut
  {NoStop}%
\bibitem [{\citenamefont {Kent}\ and\ \citenamefont
  {Zimmerman}(2025)}]{Kent:2025pvu}%
  \BibitemOpen
  \bibfield  {author} {\bibinfo {author} {\bibfnamefont {B.}~\bibnamefont
  {Kent}}\ and\ \bibinfo {author} {\bibfnamefont {A.}~\bibnamefont
  {Zimmerman}},\ }\href {\doibase 10.1103/xn1j-ddcc} {\bibfield  {journal}
  {\bibinfo  {journal} {Phys. Rev. Lett.}\ }\textbf {\bibinfo {volume} {135}},\
  \bibinfo {pages} {141501} (\bibinfo {year} {2025})},\ \Eprint
  {http://arxiv.org/abs/2505.03887} {arXiv:2505.03887 [hep-th]} \BibitemShut
  {NoStop}%
\bibitem [{\citenamefont {Chawla}\ and\ \citenamefont
  {Keeler}(2023)}]{Chawla:2023bsu}%
  \BibitemOpen
  \bibfield  {author} {\bibinfo {author} {\bibfnamefont {S.}~\bibnamefont
  {Chawla}}\ and\ \bibinfo {author} {\bibfnamefont {C.}~\bibnamefont
  {Keeler}},\ }\href {\doibase 10.1088/1361-6382/acfe57} {\bibfield  {journal}
  {\bibinfo  {journal} {Class. Quant. Grav.}\ }\textbf {\bibinfo {volume}
  {40}},\ \bibinfo {pages} {225004} (\bibinfo {year} {2023})},\ \Eprint
  {http://arxiv.org/abs/2306.02417} {arXiv:2306.02417 [hep-th]} \BibitemShut
  {NoStop}%
\bibitem [{\citenamefont {Aoude}\ and\ \citenamefont
  {Ochirov}(2023)}]{Aoude:2023fdm}%
  \BibitemOpen
  \bibfield  {author} {\bibinfo {author} {\bibfnamefont {R.}~\bibnamefont
  {Aoude}}\ and\ \bibinfo {author} {\bibfnamefont {A.}~\bibnamefont
  {Ochirov}},\ }\href {\doibase 10.1007/JHEP12(2023)103} {\bibfield  {journal}
  {\bibinfo  {journal} {JHEP}\ }\textbf {\bibinfo {volume} {12}},\ \bibinfo
  {pages} {103} (\bibinfo {year} {2023})},\ \Eprint
  {http://arxiv.org/abs/2307.07504} {arXiv:2307.07504 [hep-th]} \BibitemShut
  {NoStop}%
\bibitem [{\citenamefont {Aoude}\ \emph {et~al.}(2024)\citenamefont {Aoude},
  \citenamefont {O'Connell},\ and\ \citenamefont {Sergola}}]{Aoude:2024sve}%
  \BibitemOpen
  \bibfield  {author} {\bibinfo {author} {\bibfnamefont {R.}~\bibnamefont
  {Aoude}}, \bibinfo {author} {\bibfnamefont {D.}~\bibnamefont {O'Connell}}, \
  and\ \bibinfo {author} {\bibfnamefont {M.}~\bibnamefont {Sergola}},\
  }\href@noop {} {\  (\bibinfo {year} {2024})},\ \Eprint
  {http://arxiv.org/abs/2412.05267} {arXiv:2412.05267 [hep-th]} \BibitemShut
  {NoStop}%
\bibitem [{\citenamefont {Aoki}\ \emph {et~al.}(2025)\citenamefont {Aoki},
  \citenamefont {Cristofoli}, \citenamefont {Jeong}, \citenamefont {Sergola},\
  and\ \citenamefont {Yoshimura}}]{Aoki:2025ihc}%
  \BibitemOpen
  \bibfield  {author} {\bibinfo {author} {\bibfnamefont {K.}~\bibnamefont
  {Aoki}}, \bibinfo {author} {\bibfnamefont {A.}~\bibnamefont {Cristofoli}},
  \bibinfo {author} {\bibfnamefont {H.}~\bibnamefont {Jeong}}, \bibinfo
  {author} {\bibfnamefont {M.}~\bibnamefont {Sergola}}, \ and\ \bibinfo
  {author} {\bibfnamefont {K.}~\bibnamefont {Yoshimura}},\ }\href@noop {} {\
  (\bibinfo {year} {2025})},\ \Eprint {http://arxiv.org/abs/2509.12111}
  {arXiv:2509.12111 [hep-th]} \BibitemShut {NoStop}%
\bibitem [{\citenamefont {Affleck}\ \emph {et~al.}(1982)\citenamefont
  {Affleck}, \citenamefont {Alvarez},\ and\ \citenamefont
  {Manton}}]{Affleck:1981bma}%
  \BibitemOpen
  \bibfield  {author} {\bibinfo {author} {\bibfnamefont {I.~K.}\ \bibnamefont
  {Affleck}}, \bibinfo {author} {\bibfnamefont {O.}~\bibnamefont {Alvarez}}, \
  and\ \bibinfo {author} {\bibfnamefont {N.~S.}\ \bibnamefont {Manton}},\
  }\href {\doibase 10.1016/0550-3213(82)90455-2} {\bibfield  {journal}
  {\bibinfo  {journal} {Nucl. Phys. B}\ }\textbf {\bibinfo {volume} {197}},\
  \bibinfo {pages} {509} (\bibinfo {year} {1982})}\BibitemShut {NoStop}%
\bibitem [{\citenamefont {Bern}\ and\ \citenamefont
  {Kosower}(1991)}]{Bern:1990cu}%
  \BibitemOpen
  \bibfield  {author} {\bibinfo {author} {\bibfnamefont {Z.}~\bibnamefont
  {Bern}}\ and\ \bibinfo {author} {\bibfnamefont {D.~A.}\ \bibnamefont
  {Kosower}},\ }\href {\doibase 10.1103/PhysRevLett.66.1669} {\bibfield
  {journal} {\bibinfo  {journal} {Phys. Rev. Lett.}\ }\textbf {\bibinfo
  {volume} {66}},\ \bibinfo {pages} {1669} (\bibinfo {year}
  {1991})}\BibitemShut {NoStop}%
\bibitem [{\citenamefont {Bern}\ and\ \citenamefont
  {Kosower}(1992)}]{Bern:1991aq}%
  \BibitemOpen
  \bibfield  {author} {\bibinfo {author} {\bibfnamefont {Z.}~\bibnamefont
  {Bern}}\ and\ \bibinfo {author} {\bibfnamefont {D.~A.}\ \bibnamefont
  {Kosower}},\ }\href {\doibase 10.1016/0550-3213(92)90134-W} {\bibfield
  {journal} {\bibinfo  {journal} {Nucl. Phys. B}\ }\textbf {\bibinfo {volume}
  {379}},\ \bibinfo {pages} {451} (\bibinfo {year} {1992})}\BibitemShut
  {NoStop}%
\bibitem [{\citenamefont {Strassler}(1992)}]{Strassler:1992zr}%
  \BibitemOpen
  \bibfield  {author} {\bibinfo {author} {\bibfnamefont {M.~J.}\ \bibnamefont
  {Strassler}},\ }\href {\doibase 10.1016/0550-3213(92)90098-V} {\bibfield
  {journal} {\bibinfo  {journal} {Nucl. Phys. B}\ }\textbf {\bibinfo {volume}
  {385}},\ \bibinfo {pages} {145} (\bibinfo {year} {1992})},\ \Eprint
  {http://arxiv.org/abs/hep-ph/9205205} {arXiv:hep-ph/9205205} \BibitemShut
  {NoStop}%
\bibitem [{\citenamefont {Edwards}\ and\ \citenamefont
  {Schubert}(2019)}]{Edwards:2019eby}%
  \BibitemOpen
  \bibfield  {author} {\bibinfo {author} {\bibfnamefont {J.~P.}\ \bibnamefont
  {Edwards}}\ and\ \bibinfo {author} {\bibfnamefont {C.}~\bibnamefont
  {Schubert}}\ }(\bibinfo {year} {2019})\ \Eprint
  {http://arxiv.org/abs/1912.10004} {arXiv:1912.10004 [hep-th]} \BibitemShut
  {NoStop}%
\bibitem [{\citenamefont {K{\"a}lin}\ and\ \citenamefont
  {Porto}(2020)}]{Kalin:2020mvi}%
  \BibitemOpen
  \bibfield  {author} {\bibinfo {author} {\bibfnamefont {G.}~\bibnamefont
  {K{\"a}lin}}\ and\ \bibinfo {author} {\bibfnamefont {R.~A.}\ \bibnamefont
  {Porto}},\ }\href {\doibase 10.1007/JHEP11(2020)106} {\bibfield  {journal}
  {\bibinfo  {journal} {JHEP}\ }\textbf {\bibinfo {volume} {11}},\ \bibinfo
  {pages} {106} (\bibinfo {year} {2020})},\ \Eprint
  {http://arxiv.org/abs/2006.01184} {arXiv:2006.01184 [hep-th]} \BibitemShut
  {NoStop}%
\bibitem [{\citenamefont {Mogull}\ \emph {et~al.}(2021)\citenamefont {Mogull},
  \citenamefont {Plefka},\ and\ \citenamefont {Steinhoff}}]{Mogull:2020sak}%
  \BibitemOpen
  \bibfield  {author} {\bibinfo {author} {\bibfnamefont {G.}~\bibnamefont
  {Mogull}}, \bibinfo {author} {\bibfnamefont {J.}~\bibnamefont {Plefka}}, \
  and\ \bibinfo {author} {\bibfnamefont {J.}~\bibnamefont {Steinhoff}},\ }\href
  {\doibase 10.1007/JHEP02(2021)048} {\bibfield  {journal} {\bibinfo  {journal}
  {JHEP}\ }\textbf {\bibinfo {volume} {02}},\ \bibinfo {pages} {048} (\bibinfo
  {year} {2021})},\ \Eprint {http://arxiv.org/abs/2010.02865} {arXiv:2010.02865
  [hep-th]} \BibitemShut {NoStop}%
\bibitem [{\citenamefont {Gonzo}\ and\ \citenamefont
  {Shi}(2021)}]{Gonzo:2021drq}%
  \BibitemOpen
  \bibfield  {author} {\bibinfo {author} {\bibfnamefont {R.}~\bibnamefont
  {Gonzo}}\ and\ \bibinfo {author} {\bibfnamefont {C.}~\bibnamefont {Shi}},\
  }\href {\doibase 10.1103/PhysRevD.104.105012} {\bibfield  {journal} {\bibinfo
   {journal} {Phys. Rev. D}\ }\textbf {\bibinfo {volume} {104}},\ \bibinfo
  {pages} {105012} (\bibinfo {year} {2021})},\ \Eprint
  {http://arxiv.org/abs/2109.01072} {arXiv:2109.01072 [hep-th]} \BibitemShut
  {NoStop}%
\bibitem [{\citenamefont {Hartle}\ and\ \citenamefont
  {Hawking}(1976)}]{Hartle:1976tp}%
  \BibitemOpen
  \bibfield  {author} {\bibinfo {author} {\bibfnamefont {J.~B.}\ \bibnamefont
  {Hartle}}\ and\ \bibinfo {author} {\bibfnamefont {S.~W.}\ \bibnamefont
  {Hawking}},\ }\href {\doibase 10.1103/PhysRevD.13.2188} {\bibfield  {journal}
  {\bibinfo  {journal} {Phys. Rev. D}\ }\textbf {\bibinfo {volume} {13}},\
  \bibinfo {pages} {2188} (\bibinfo {year} {1976})}\BibitemShut {NoStop}%
\bibitem [{\citenamefont {Chitre}\ and\ \citenamefont
  {Hartle}(1977)}]{Chitre:1977ip}%
  \BibitemOpen
  \bibfield  {author} {\bibinfo {author} {\bibfnamefont {D.~M.}\ \bibnamefont
  {Chitre}}\ and\ \bibinfo {author} {\bibfnamefont {J.~B.}\ \bibnamefont
  {Hartle}},\ }\href {\doibase 10.1103/PhysRevD.16.251} {\bibfield  {journal}
  {\bibinfo  {journal} {Phys. Rev. D}\ }\textbf {\bibinfo {volume} {16}},\
  \bibinfo {pages} {251} (\bibinfo {year} {1977})}\BibitemShut {NoStop}%
\bibitem [{\citenamefont {Srinivasan}\ and\ \citenamefont
  {Padmanabhan}(1999)}]{Srinivasan:1998ty}%
  \BibitemOpen
  \bibfield  {author} {\bibinfo {author} {\bibfnamefont {K.}~\bibnamefont
  {Srinivasan}}\ and\ \bibinfo {author} {\bibfnamefont {T.}~\bibnamefont
  {Padmanabhan}},\ }\href {\doibase 10.1103/PhysRevD.60.024007} {\bibfield
  {journal} {\bibinfo  {journal} {Phys. Rev. D}\ }\textbf {\bibinfo {volume}
  {60}},\ \bibinfo {pages} {024007} (\bibinfo {year} {1999})},\ \Eprint
  {http://arxiv.org/abs/gr-qc/9812028} {arXiv:gr-qc/9812028} \BibitemShut
  {NoStop}%
\bibitem [{\citenamefont {Semr{\'e}n}\ and\ \citenamefont
  {Torgrimsson}(2025)}]{Semren:2025dix}%
  \BibitemOpen
  \bibfield  {author} {\bibinfo {author} {\bibfnamefont {P.}~\bibnamefont
  {Semr{\'e}n}}\ and\ \bibinfo {author} {\bibfnamefont {G.}~\bibnamefont
  {Torgrimsson}},\ }\href@noop {} {\  (\bibinfo {year} {2025})},\ \Eprint
  {http://arxiv.org/abs/2508.01901} {arXiv:2508.01901 [hep-th]} \BibitemShut
  {NoStop}%
\bibitem [{\citenamefont {Ilderton}\ and\ \citenamefont
  {Rajeev}(2025)}]{Ilderton:2025umd}%
  \BibitemOpen
  \bibfield  {author} {\bibinfo {author} {\bibfnamefont {A.}~\bibnamefont
  {Ilderton}}\ and\ \bibinfo {author} {\bibfnamefont {K.}~\bibnamefont
  {Rajeev}},\ }\href@noop {} {\  (\bibinfo {year} {2025})},\ \Eprint
  {http://arxiv.org/abs/2508.00997} {arXiv:2508.00997 [hep-th]} \BibitemShut
  {NoStop}%
\bibitem [{\citenamefont {Vaidya}(1966)}]{Vaidya:1966zza}%
  \BibitemOpen
  \bibfield  {author} {\bibinfo {author} {\bibfnamefont {P.~C.}\ \bibnamefont
  {Vaidya}},\ }\href {\doibase 10.1086/148692} {\bibfield  {journal} {\bibinfo
  {journal} {Astrophys. J.}\ }\textbf {\bibinfo {volume} {144}},\ \bibinfo
  {pages} {943} (\bibinfo {year} {1966})}\BibitemShut {NoStop}%
\bibitem [{\citenamefont {Arkani-Hamed}\ \emph {et~al.}(2020)\citenamefont
  {Arkani-Hamed}, \citenamefont {Huang},\ and\ \citenamefont
  {O'Connell}}]{Arkani-Hamed:2019ymq}%
  \BibitemOpen
  \bibfield  {author} {\bibinfo {author} {\bibfnamefont {N.}~\bibnamefont
  {Arkani-Hamed}}, \bibinfo {author} {\bibfnamefont {Y.-t.}\ \bibnamefont
  {Huang}}, \ and\ \bibinfo {author} {\bibfnamefont {D.}~\bibnamefont
  {O'Connell}},\ }\href {\doibase 10.1007/JHEP01(2020)046} {\bibfield
  {journal} {\bibinfo  {journal} {JHEP}\ }\textbf {\bibinfo {volume} {01}},\
  \bibinfo {pages} {046} (\bibinfo {year} {2020})},\ \Eprint
  {http://arxiv.org/abs/1906.10100} {arXiv:1906.10100 [hep-th]} \BibitemShut
  {NoStop}%
\bibitem [{\citenamefont {Guevara}\ \emph {et~al.}(2021)\citenamefont
  {Guevara}, \citenamefont {Maybee}, \citenamefont {Ochirov}, \citenamefont
  {O'Connell},\ and\ \citenamefont {Vines}}]{Guevara:2020xjx}%
  \BibitemOpen
  \bibfield  {author} {\bibinfo {author} {\bibfnamefont {A.}~\bibnamefont
  {Guevara}}, \bibinfo {author} {\bibfnamefont {B.}~\bibnamefont {Maybee}},
  \bibinfo {author} {\bibfnamefont {A.}~\bibnamefont {Ochirov}}, \bibinfo
  {author} {\bibfnamefont {D.}~\bibnamefont {O'Connell}}, \ and\ \bibinfo
  {author} {\bibfnamefont {J.}~\bibnamefont {Vines}},\ }\href {\doibase
  10.1007/JHEP03(2021)201} {\bibfield  {journal} {\bibinfo  {journal} {JHEP}\
  }\textbf {\bibinfo {volume} {03}},\ \bibinfo {pages} {201} (\bibinfo {year}
  {2021})},\ \Eprint {http://arxiv.org/abs/2012.11570} {arXiv:2012.11570
  [hep-th]} \BibitemShut {NoStop}%
\bibitem [{\citenamefont {Goldberger}\ and\ \citenamefont
  {Ridgway}(2017)}]{Goldberger:2016iau}%
  \BibitemOpen
  \bibfield  {author} {\bibinfo {author} {\bibfnamefont {W.~D.}\ \bibnamefont
  {Goldberger}}\ and\ \bibinfo {author} {\bibfnamefont {A.~K.}\ \bibnamefont
  {Ridgway}},\ }\href {\doibase 10.1103/PhysRevD.95.125010} {\bibfield
  {journal} {\bibinfo  {journal} {Phys. Rev. D}\ }\textbf {\bibinfo {volume}
  {95}},\ \bibinfo {pages} {125010} (\bibinfo {year} {2017})},\ \Eprint
  {http://arxiv.org/abs/1611.03493} {arXiv:1611.03493 [hep-th]} \BibitemShut
  {NoStop}%
\bibitem [{\citenamefont {Moynihan}\ \emph {et~al.}(2025)\citenamefont
  {Moynihan}, \citenamefont {Ashby},\ and\ \citenamefont
  {White}}]{Moynihan:2025vcs}%
  \BibitemOpen
  \bibfield  {author} {\bibinfo {author} {\bibfnamefont {N.}~\bibnamefont
  {Moynihan}}, \bibinfo {author} {\bibfnamefont {M.~L.~R.}\ \bibnamefont
  {Ashby}}, \ and\ \bibinfo {author} {\bibfnamefont {C.~D.}\ \bibnamefont
  {White}},\ }\href@noop {} {\  (\bibinfo {year} {2025})},\ \Eprint
  {http://arxiv.org/abs/2509.22350} {arXiv:2509.22350 [hep-th]} \BibitemShut
  {NoStop}%
\bibitem [{Note1()}]{Note1}%
  \BibitemOpen
  \bibinfo {note} {Our conventions are such that what we call a `particle' is
  attracted to the source, in both Yang Mills and gravity.}\BibitemShut {Stop}%
\bibitem [{\citenamefont {Di~Vecchia}\ \emph {et~al.}(2024)\citenamefont
  {Di~Vecchia}, \citenamefont {Heissenberg}, \citenamefont {Russo},\ and\
  \citenamefont {Veneziano}}]{DiVecchia:2023frv}%
  \BibitemOpen
  \bibfield  {author} {\bibinfo {author} {\bibfnamefont {P.}~\bibnamefont
  {Di~Vecchia}}, \bibinfo {author} {\bibfnamefont {C.}~\bibnamefont
  {Heissenberg}}, \bibinfo {author} {\bibfnamefont {R.}~\bibnamefont {Russo}},
  \ and\ \bibinfo {author} {\bibfnamefont {G.}~\bibnamefont {Veneziano}},\
  }\href {\doibase 10.1016/j.physrep.2024.06.002} {\bibfield  {journal}
  {\bibinfo  {journal} {Phys. Rept.}\ }\textbf {\bibinfo {volume} {1083}},\
  \bibinfo {pages} {1} (\bibinfo {year} {2024})},\ \Eprint
  {http://arxiv.org/abs/2306.16488} {arXiv:2306.16488 [hep-th]} \BibitemShut
  {NoStop}%
\bibitem [{\citenamefont {Aoude}\ \emph {et~al.}(2025)\citenamefont {Aoude},
  \citenamefont {O'Connell}, \citenamefont {Sergola},\ and\ \citenamefont
  {White}}]{Aoude:2025jvt}%
  \BibitemOpen
  \bibfield  {author} {\bibinfo {author} {\bibfnamefont {R.}~\bibnamefont
  {Aoude}}, \bibinfo {author} {\bibfnamefont {D.}~\bibnamefont {O'Connell}},
  \bibinfo {author} {\bibfnamefont {M.}~\bibnamefont {Sergola}}, \ and\
  \bibinfo {author} {\bibfnamefont {C.~D.}\ \bibnamefont {White}},\ }\href@noop
  {} {\  (\bibinfo {year} {2025})},\ \Eprint {http://arxiv.org/abs/2510.25866}
  {arXiv:2510.25866 [hep-th]} \BibitemShut {NoStop}%
\bibitem [{\citenamefont {Gerlach}(1976)}]{Gerlach:1976ji}%
  \BibitemOpen
  \bibfield  {author} {\bibinfo {author} {\bibfnamefont {U.~H.}\ \bibnamefont
  {Gerlach}},\ }\href {\doibase 10.1103/PhysRevD.14.1479} {\bibfield  {journal}
  {\bibinfo  {journal} {Phys. Rev. D}\ }\textbf {\bibinfo {volume} {14}},\
  \bibinfo {pages} {1479} (\bibinfo {year} {1976})}\BibitemShut {NoStop}%
\bibitem [{\citenamefont {Vachaspati}\ \emph {et~al.}(2007)\citenamefont
  {Vachaspati}, \citenamefont {Stojkovic},\ and\ \citenamefont
  {Krauss}}]{Vachaspati:2006ki}%
  \BibitemOpen
  \bibfield  {author} {\bibinfo {author} {\bibfnamefont {T.}~\bibnamefont
  {Vachaspati}}, \bibinfo {author} {\bibfnamefont {D.}~\bibnamefont
  {Stojkovic}}, \ and\ \bibinfo {author} {\bibfnamefont {L.~M.}\ \bibnamefont
  {Krauss}},\ }\href {\doibase 10.1103/PhysRevD.76.024005} {\bibfield
  {journal} {\bibinfo  {journal} {Phys. Rev. D}\ }\textbf {\bibinfo {volume}
  {76}},\ \bibinfo {pages} {024005} (\bibinfo {year} {2007})},\ \Eprint
  {http://arxiv.org/abs/gr-qc/0609024} {arXiv:gr-qc/0609024} \BibitemShut
  {NoStop}%
\bibitem [{\citenamefont {Fradkin}\ \emph {et~al.}(1991)\citenamefont
  {Fradkin}, \citenamefont {Gitman},\ and\ \citenamefont
  {Shvartsman}}]{Fradkin1991QED}%
  \BibitemOpen
  \bibfield  {author} {\bibinfo {author} {\bibfnamefont {E.~S.}\ \bibnamefont
  {Fradkin}}, \bibinfo {author} {\bibfnamefont {D.~M.}\ \bibnamefont {Gitman}},
  \ and\ \bibinfo {author} {\bibfnamefont {S.~M.}\ \bibnamefont {Shvartsman}},\
  }\href@noop {} {\emph {\bibinfo {title} {Quantum Electrodynamics with
  Unstable Vacuum}}},\ Springer Series in Nuclear and Particle Physics\
  (\bibinfo  {publisher} {Springer-Verlag},\ \bibinfo {address} {Berlin},\
  \bibinfo {year} {1991})\BibitemShut {NoStop}%
\bibitem [{\citenamefont {Wald}(1994)}]{Wald1994QFT}%
  \BibitemOpen
  \bibfield  {author} {\bibinfo {author} {\bibfnamefont {R.~M.}\ \bibnamefont
  {Wald}},\ }\href@noop {} {\emph {\bibinfo {title} {Quantum Field Theory in
  Curved Spacetime and Black Hole Thermodynamics}}},\ Chicago Lectures in
  Physics\ (\bibinfo  {publisher} {University of Chicago Press},\ \bibinfo
  {address} {Chicago},\ \bibinfo {year} {1994})\BibitemShut {NoStop}%
\bibitem [{\citenamefont {Feldbrugge}(2019)}]{Feldbrugge:2019sew}%
  \BibitemOpen
  \bibfield  {author} {\bibinfo {author} {\bibfnamefont {J.~L.}\ \bibnamefont
  {Feldbrugge}},\ }\emph {\bibinfo {title} {{Path Integrals in the Sky:
  Classical and Quantum Problems with Minimal Assumptions}}},\ \href@noop {}
  {Ph.D. thesis},\ \bibinfo  {school} {U. Waterloo (main)} (\bibinfo {year}
  {2019})\BibitemShut {NoStop}%
\bibitem [{\citenamefont {Rajeev}(2021)}]{Rajeev:2021zae}%
  \BibitemOpen
  \bibfield  {author} {\bibinfo {author} {\bibfnamefont {K.}~\bibnamefont
  {Rajeev}},\ }\href {\doibase 10.1103/PhysRevD.104.105014} {\bibfield
  {journal} {\bibinfo  {journal} {Phys. Rev. D}\ }\textbf {\bibinfo {volume}
  {104}},\ \bibinfo {pages} {105014} (\bibinfo {year} {2021})},\ \Eprint
  {http://arxiv.org/abs/2105.12194} {arXiv:2105.12194 [hep-th]} \BibitemShut
  {NoStop}%
\bibitem [{\citenamefont {Hawking}(1975)}]{Hawking:1975vcx}%
  \BibitemOpen
  \bibfield  {author} {\bibinfo {author} {\bibfnamefont {S.~W.}\ \bibnamefont
  {Hawking}},\ }\href {\doibase 10.1007/BF02345020} {\bibfield  {journal}
  {\bibinfo  {journal} {Commun. Math. Phys.}\ }\textbf {\bibinfo {volume}
  {43}},\ \bibinfo {pages} {199} (\bibinfo {year} {1975})},\ \bibinfo {note}
  {[Erratum: Commun.Math.Phys. 46, 206 (1976)]}\BibitemShut {NoStop}%
\bibitem [{Note2()}]{Note2}%
  \BibitemOpen
  \bibinfo {note} {{We note in passing that the prescription $qQ\rightarrow
  M\protect \mathcal {E}$ mentioned ahead of~\protect \textup {\hbox
  {\mathsurround \z@ \protect \normalfont (\ignorespaces \ref
  {new-DC-rule-1}\unskip \@@italiccorr )}}}, enforced at the level of the
  \protect \emph {action} by $qQ\rightarrow -M\partial _{v} S_{cl}(v)$ {where}
  $\partial _{v}S_{cl}(v)=-\protect \mathcal {E}'+\Delta \protect \mathcal
  {E}$, also gives the correct semiclassical action relevant to the mode
  function~(\ref {eq:out-gen}). It would be interesting to explore the
  connection between this procedure and the observation in~\cite
  {Adamo:2022rmp} that certain aspects of double copy in plane-wave backgrounds
  manifest when momentum conserving delta-functions on the gauge-theory side
  are represented in position-space form.}\BibitemShut {Stop}%
\bibitem [{\citenamefont {Adamo}\ \emph {et~al.}(2024)\citenamefont {Adamo},
  \citenamefont {Cristofoli}, \citenamefont {Ilderton},\ and\ \citenamefont
  {Klisch}}]{Adamo:2023cfp}%
  \BibitemOpen
  \bibfield  {author} {\bibinfo {author} {\bibfnamefont {T.}~\bibnamefont
  {Adamo}}, \bibinfo {author} {\bibfnamefont {A.}~\bibnamefont {Cristofoli}},
  \bibinfo {author} {\bibfnamefont {A.}~\bibnamefont {Ilderton}}, \ and\
  \bibinfo {author} {\bibfnamefont {S.}~\bibnamefont {Klisch}},\ }\href
  {\doibase 10.1088/1361-6382/ad210f} {\bibfield  {journal} {\bibinfo
  {journal} {Class. Quant. Grav.}\ }\textbf {\bibinfo {volume} {41}},\ \bibinfo
  {pages} {065006} (\bibinfo {year} {2024})},\ \Eprint
  {http://arxiv.org/abs/2307.00431} {arXiv:2307.00431 [hep-th]} \BibitemShut
  {NoStop}%
\bibitem [{\citenamefont {Hawking}(1974)}]{Hawking:1974rv}%
  \BibitemOpen
  \bibfield  {author} {\bibinfo {author} {\bibfnamefont {S.~W.}\ \bibnamefont
  {Hawking}},\ }\href {\doibase 10.1038/248030a0} {\bibfield  {journal}
  {\bibinfo  {journal} {Nature}\ }\textbf {\bibinfo {volume} {248}},\ \bibinfo
  {pages} {30} (\bibinfo {year} {1974})}\BibitemShut {NoStop}%
\bibitem [{\citenamefont {{Digital Library of Mathematical
  Functions}}(1314)}]{DLMF:Whittaker}%
  \BibitemOpen
  \bibfield  {author} {\bibinfo {author} {\bibnamefont {{Digital Library of
  Mathematical Functions}}},\ }\href@noop {} {\bibfield  {journal} {\bibinfo
  {journal} {{Section 13.14}}\ } (\bibinfo {year}
  {https://dlmf.nist.gov/13.14})}\BibitemShut {NoStop}%
\bibitem [{\citenamefont {Adamo}\ \emph
  {et~al.}(2022{\natexlab{b}})\citenamefont {Adamo}, \citenamefont
  {Cristofoli},\ and\ \citenamefont {Ilderton}}]{Adamo:2022rmp}%
  \BibitemOpen
  \bibfield  {author} {\bibinfo {author} {\bibfnamefont {T.}~\bibnamefont
  {Adamo}}, \bibinfo {author} {\bibfnamefont {A.}~\bibnamefont {Cristofoli}}, \
  and\ \bibinfo {author} {\bibfnamefont {A.}~\bibnamefont {Ilderton}},\ }\href
  {\doibase 10.1007/JHEP08(2022)281} {\bibfield  {journal} {\bibinfo  {journal}
  {JHEP}\ }\textbf {\bibinfo {volume} {08}},\ \bibinfo {pages} {281} (\bibinfo
  {year} {2022}{\natexlab{b}})},\ \Eprint {http://arxiv.org/abs/2203.13785}
  {arXiv:2203.13785 [hep-th]} \BibitemShut {NoStop}%
\end{thebibliography}
\end{document}